\documentclass[preprint]{aastex}

\def\<#1>{[\hbox{#1}]}
\def\hoverarrow#1{\setbox0\hbox to 0pt{\hss$\scriptstyle\rightarrow$}%
                 #1\kern.6ex\raise 1.6ex\box0\kern-.2ex}
\def\btheta{{\hoverarrow\theta}}
\def\bbeta{{\hoverarrow\beta}}
\def\overarrow#1{\setbox0\hbox to 0pt{\hss$\scriptstyle\rightarrow$}%
                 #1\raise 1.6ex\box0}
\def\bnabla{{\overarrow\nabla}}
\def\({\raise.3ex\hbox{(}} \def\){\raise.3ex\hbox{)}}

\def\half{\hbox{$1\over2$}}

\def\Tzls{T(z_{\rm L},z_{\rm S})}
\def\kann{\langle\kappa\rangle_{\rm ann}}

\def\begintype{$$\vbox\bgroup\leftskip=2\parindent \parindent=0pt\tt}
\def\endtype{\egroup$$}
\def\+{\phantom+}

\begin{document}

\title{A portable modeler of lensed quasars}

       \author{Prasenjit Saha}
       \affil{Astronomy Unit\\
              Queen Mary and Westfield College\\
              University of London\\
              London E1~4NS, UK}
       \email{p.saha@qmul.ac.uk}

       \and

       \author{Liliya L.R. Williams}
       \affil{Department of Astronomy\\
              University of Minnesota\\
	      116 Church Street SE\\
              Minneapolis, MN 55455}
       \email{llrw@astro.umn.edu}

\begin{abstract}
We introduce and implement two novel ideas for modeling lensed
quasars.  The first idea is to require different lenses to agree about
$H_0$.  This means that some models for one lens can be ruled out by
data on a different lens.  We explain using two worked examples.  One
example models 1115+080, 1608+656 (time-delay quads)
and 1933+503 (a prospective time-delay system)
all together, yielding time-delay predictions for the third lens
and a 90\%-confidence estimate of
${H_0}^{-1}=14.6_{-1.7}^{+9.4}$~Gyr
($H_0=67_{-26}^{\,+\;9}\rm\;\,km\;s^{-1}\,Mpc^{-1}$)
assuming $(\Omega_M=0.3,\Omega_\Lambda=0.7)$.
The other example models the time-delay doubles
1520+530, 1600+434, 1830-211, and 2149-275, which gives
${H_0}^{-1}=14.5_{-1.5}^{+3.3}$~Gyr
($H_0=67_{-13}^{\,+\;8}\rm\;\,km\;s^{-1}\,Mpc^{-1}$).
Our second idea is to write the whole modeling software as a highly
interactive Java applet, which can be used both for coarse-grained
results inside a browser and for fine-grained results on a
workstation.  Several obstacles come up in trying to implement a
numerically-intensive method thus, but we overcome them.
\end{abstract}


\section{Introduction}

Some aspects of modeling lensed quasars are much as they were just
after the discovery of the first double quasar 0957+561.  In one of
the earliest lens-modeling papers, \cite{young81} are concerned with
some now-very-familiar issues: the effect on image positions of both
the main lensing galaxy and other galaxies, the time delays predicted
by the models, the non-uniqueness of the models despite the adequacy
of the data, and the desirability of supplementary data about the
lens, such as velocity dispersions or X-rays.

But other aspects of the subject these days would have been
unimaginable in 1981.  The first double quasar has been joined by
dozens of others: the CASTLES survey compilation \citep{castles}
currently lists 76 secure multiple-image galaxy-lens systems, with
image positions measured to the mas level, even more precisely if
there are compact radio sources.  And the time delay for 0957+561, for
which Young et al.'s preliminary estimate was about 5 years, is now
measured as $423\pm1\,\rm days$ \citep{oscoz01}, along with time-delay
measurements for eight other systems
\citep{schechter97,lovell98,biggs99,burud00,cohen00,%
burud02a,burud02b,fassnacht02,hjorth02}.

These excellent data demand new, more automatic, and more portable
software tools for modeling the lenses.  Thus motivated, we have
developed a new code, {\em PixeLens,} which we present in this paper.

{\em PixeLens\/} works by reconstructing a pixelated mass map for the
lens, an idea we first implemented in \cite{sw97}.  Most lens modeling
codes work by fitting a parametric functional form---for example,
\cite{young81} fitted King models; {\em gravlens\/} by \cite{keeton01}
is a modern example, offering the user a large choice of parametric models.
There are other possibilities too:
\cite{trotter00} reconstruct lenses non-parametrically as we do, but
use multipole expansions rather than pixels.

{\em PixeLens\/} generates large ensembles of models rather than
one or a few mass maps, as a way of addressing the non-uniqueness
problem.  We introduced this strategy in \cite{ws00} for pixelated
models.  \cite{kw03} use a somewhat similar strategy, but with
parametrized models.

But {\em PixeLens\/} also brings two completely new features, one
astrophysical, one computational:
\begin{itemize}
\item It can model several lenses simultaneously, enforcing
consistency of $H_0$ across different time-delay lenses. As a result,
lenses can be used to constrain other lenses in an interesting and
surprising way;
\item The code is highly portable and can run without change as a
standalone program, or as a Java applet inside a web browser.  In the
online version of this paper, it is available as an alternative
version of Figure~\ref{gui}.
\end{itemize}

The best way to explain what we have implemented and what problems
remain is through an example.  So in the following section we will
work through a simultaneous reconstruction of three lenses: the time
delay quads 1115+080 and 1608+656, and the ten-image system 1933+503.
This will be the main part of the paper.  We turn to some
computational issues after that.

\section{A worked example: reconstructing three lenses}

In our first worked example, we consider the three lenses currently
best-constrained by observations.
\begin{enumerate}
\item 1115+080 is a quad where the lens is an elliptical galaxy
supplemented by external shear from other group galaxies.  It was the
second lens to be discovered \citep{weymann80}.  \cite{schechter97},
together with an improved time-series analysis by \cite{barkana97}
provide time-delays between two pairs of images.
\item 1608+656 is a quad with the lens being apparently two
interacting galaxies.  It was discovered \citep{myers95} in the CLASS
survey.  There are time delays between all pairs of images
\citep{fassnacht02}.
\item 1933+503 is really two quads and a double.  The lens appears to
be an elliptical, but the source is a radio source with a core and
opposing lobes.  The core and one of the lobes are imaged as quads
while the other lobe is imaged as a double.  It was also discovered in
the CLASS survey \citep{sykes98}.  There are no measured time-delays,
but there are prospects for time delays from the core quad.  We will
predict time delays for the core quad.
\end{enumerate}
The CASTLES web page provides further information.
We can also make some preliminary inferences about the lenses by
examining the morphology, as explained in \cite{sw03}: we can work
out the time-ordering of each of the image systems, and for 1115+080
we can recognize an external shear and its approximate axis.

\begin{figure}
\epsscale{0.6} 
\plotone{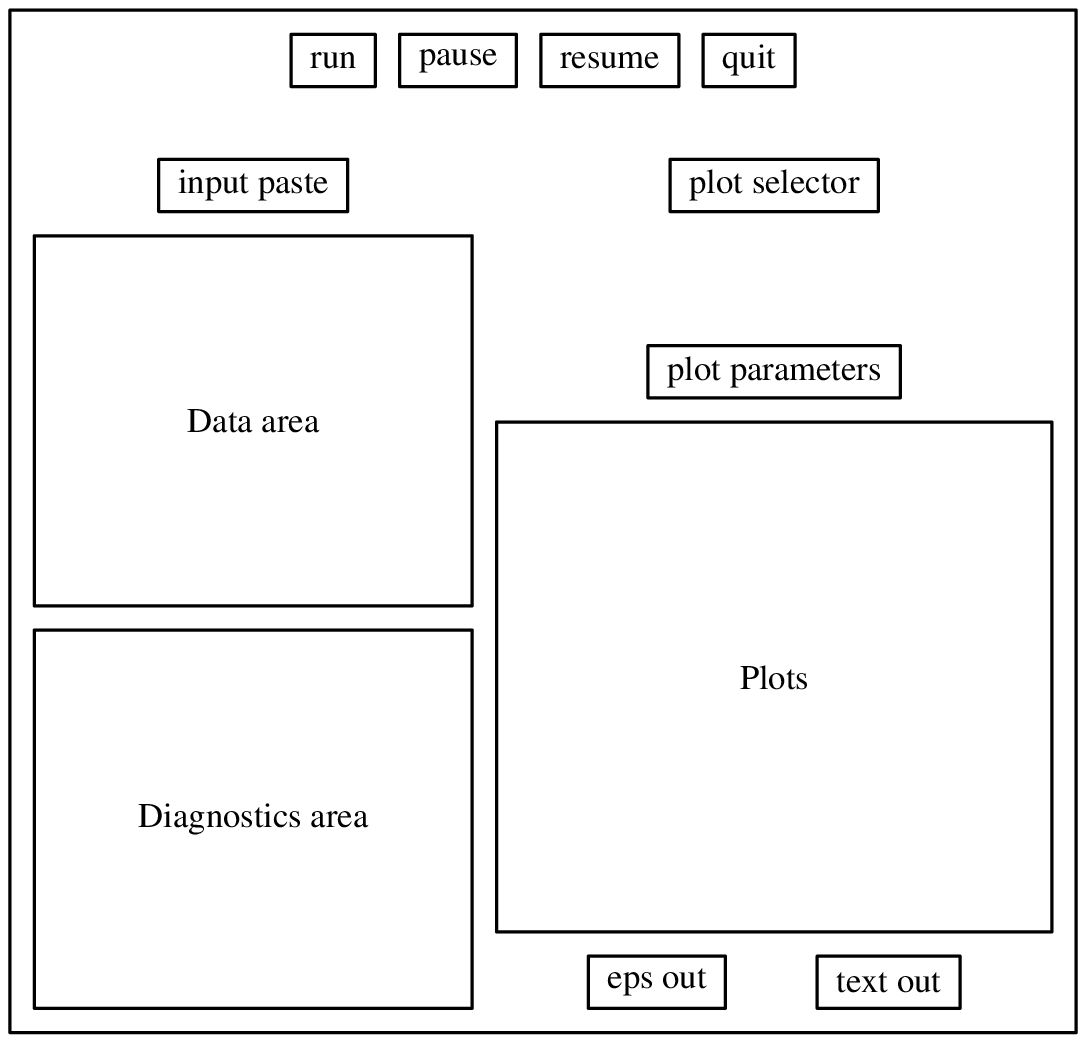}
\caption{The appearance of the {\em PixeLens\/} user interface, with all
the areas marked.}
\label{gui}
\end{figure}

Let us now proceed to apply {\em PixeLens\/}.  Figure \ref{gui} shows
the user interface, which we will explain in stages below.

\subsection{The data input}

Data is entered by typing into the `data area'.  (There is also an
`input~paste' feature with a number of example inputs, which can be
pasted and then edited by selecting the `edit' option of the menu.)

For 1115+080, we enter the following.

\begintype
object 1115               \hfil\break
symm pixrad 12 maprad 2   \hfil\break
redshifts 0.311 1.722     \hfil\break
shear -45                 \hfil\break
quad                      \hfil\break
\+0.3550 \+1.3220         \hfil\break
-0.9090 -0.7140 13.3      \hfil\break
-1.0930 -0.2600 0         \hfil\break
\+0.7170 -0.6270 11.7
\endtype
The first line just labels the object. The second line specifies that
the mass maps be inversion-symmetric with a radius of $2''$ and 12
pixels (thus setting the pixel size).  The fourth line gives the
approximate shear axis (`{\tt shear 135}' would be equivalent), in
degrees relative to East.  The program will consider positions angles
within $45^\circ$ of what is given.  The last four lines give the
image positions (in arcseconds and relative to the lens center) and
the time delays (in days and relative to the previous image).  The
images must be ordered by arrival-time.  The `{\tt 0}' in the
second-last line means that the time-delay is $>0$ but unknown.
Otherwise, uncertainties in image positions and time delays are
assumed negligible.

Note how the input format assumes that some preliminary analysis, of
the qualitative variety explained in \cite{sw03} has already been
done.

For 1608+656 we enter the following.
\begintype
object 1608               \hfil\break
pixrad 9 maprad 2         \hfil\break
redshifts 0.630 1.394     \hfil\break
quad                      \hfil\break
 -1.300  -0.800           \hfil\break
 -0.560 \+1.160   31      \hfil\break
 -1.310 \+0.700  \+5      \hfil\break
\+0.570  -0.080   40
\endtype
The format is just as before, except that we have left out {\tt symm},
because the lens here appears to be asymmetric.

For 1933+503 we enter the following.
\begintype
object 1933                           \hfil\break
symm pixrad 10                        \hfil\break
redshifts 0.76 2.63                   \hfil\break
shear 45                              \hfil\break
quad  -0.40 \+0.53 \+0.43 -0.26 0     \hfil\break
\+0.40 \+0.19 0  -0.19  -0.33 0       \hfil\break
quad \+0.54 -0.44  -0.15 \+0.44 0     \hfil\break
 -0.03 \+0.45 0  -0.38  -0.12 0       \hfil\break
double -0.47 \+0.60 \+0.13 -0.30 0
\endtype
Here we have left out the mass-map radius, leaving the program to
choose it.  We have also given image data on two quads and a double
rather than just one quad, and indicated unknown time delays with
many {\tt0}s.  Line breaks in the input have no significance.

\begin{figure}
\epsscale{0.38}
\plotone{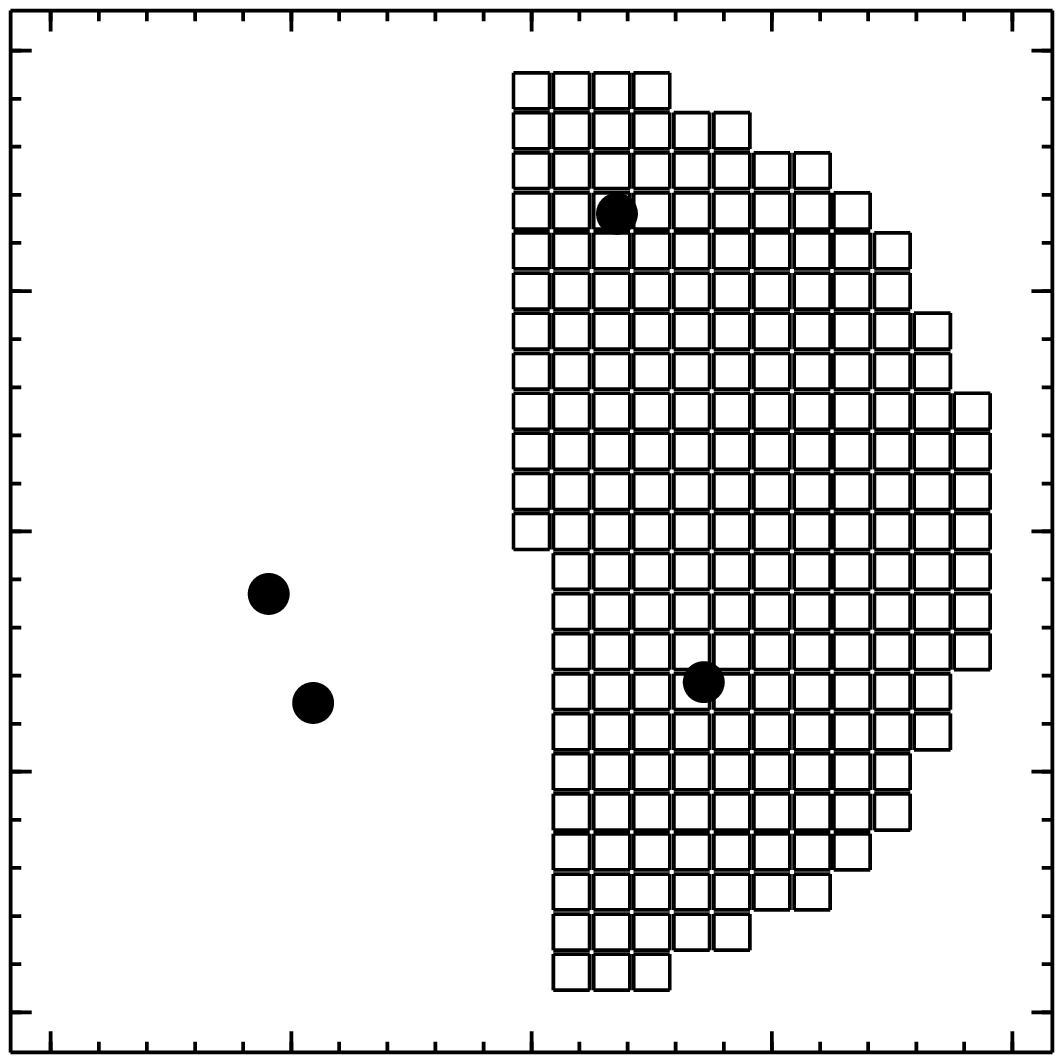}\par
\plotone{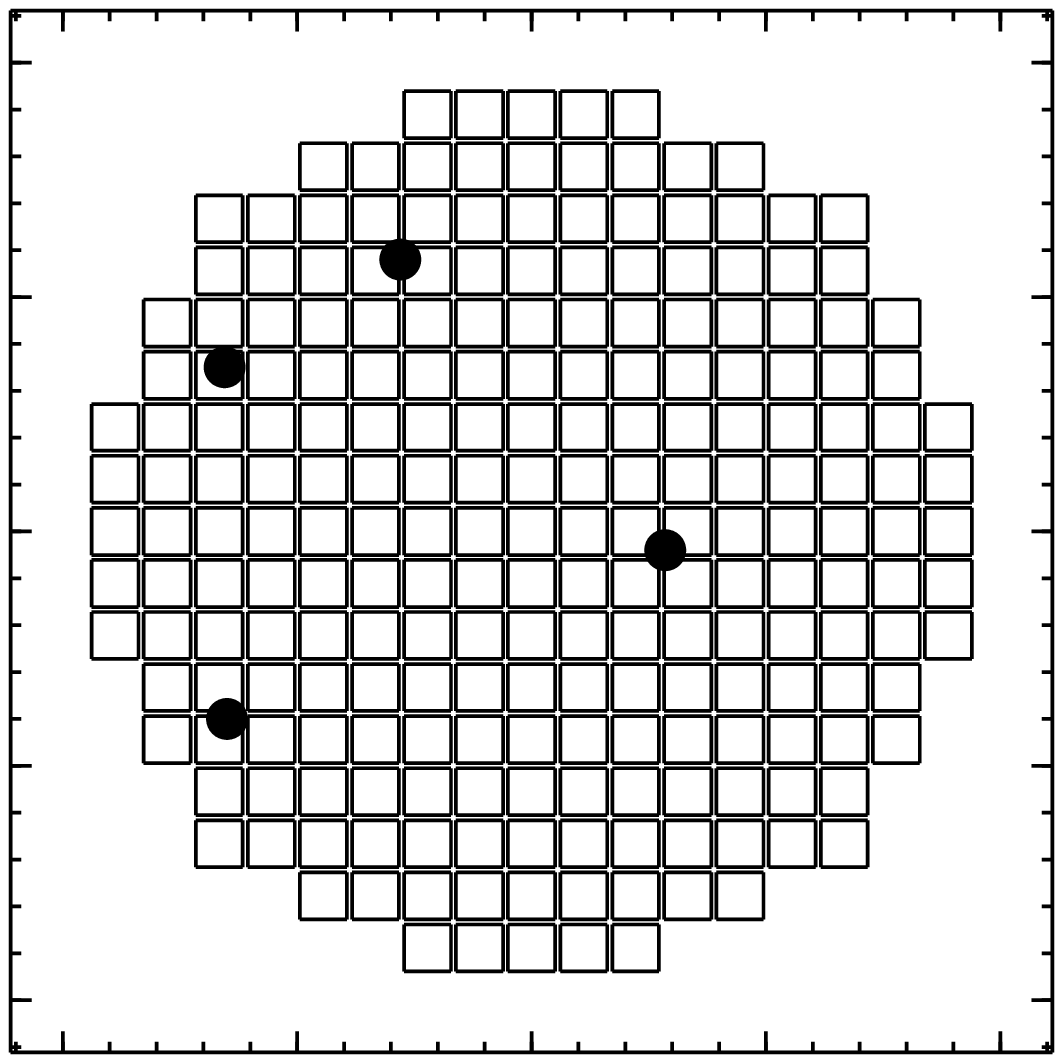}\par
\plotone{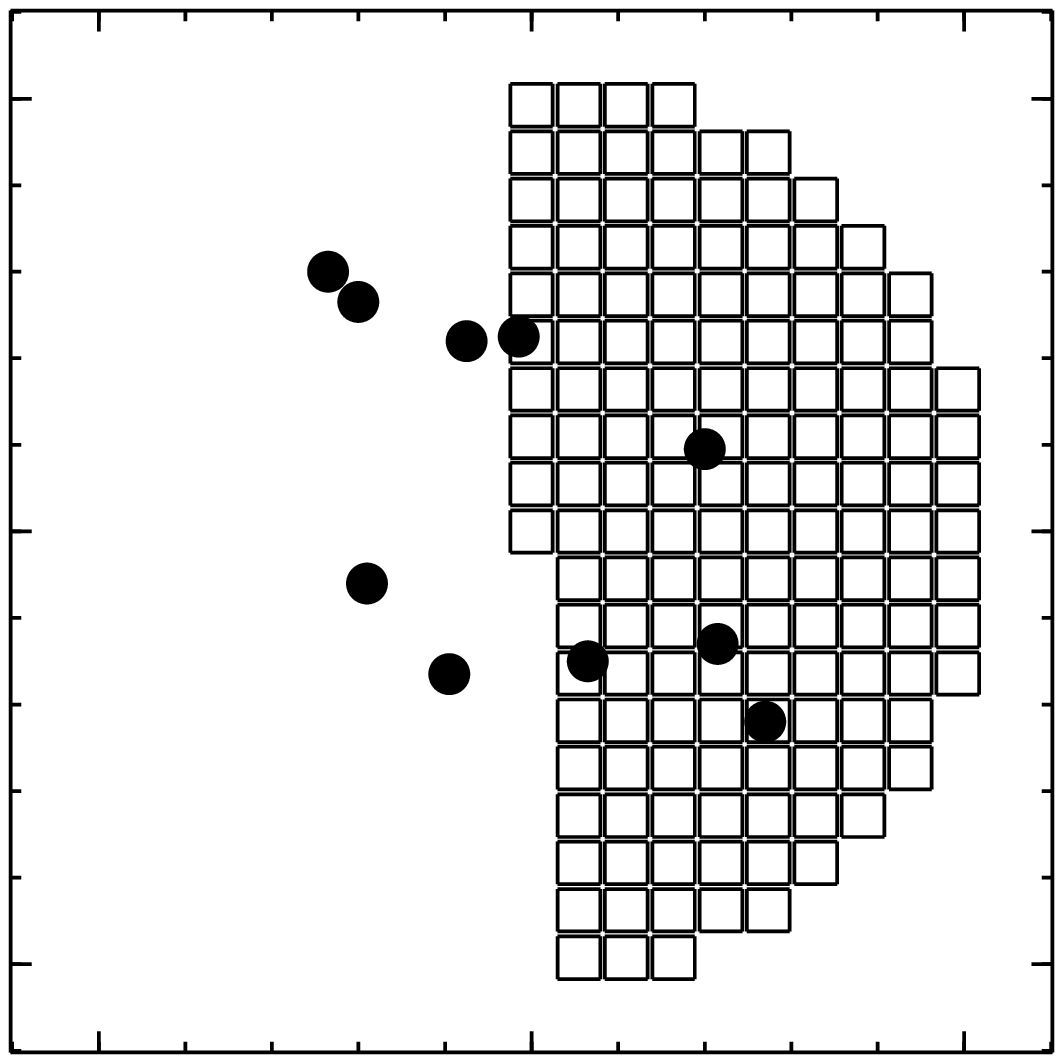}
\caption{Image positions and pixelation for the three lens systems:
1115+080 (upper), 1608+656 (middle), and 1933+503 (lower).}
\label{pix}
\end{figure}

Figure \ref{pix} summarizes the image data and the pixelation.  (For
symmetric mass maps, only half the pixels are shown here, the other
half are specified by inversion symmetry.)  {\em PixeLens\/} produces
such plots before starting the real computations, and they provide a
convenient goof test on the input.

Now we type
\begintype
models 100
\endtype
in the `data area' and click on the `run' button, and away we go.

\subsection{Generating an ensemble of models}\label{prior-sec}

At this point, we recall some lensing theory.  For a source at
$\bbeta$, the arrival-time at $\btheta$ is
\begin{equation}
\tau\(\btheta\) = \half|\btheta|^2 - \btheta\!\cdot\!\bbeta
- \int\! \ln|\btheta-\btheta{}'| \, \kappa\(\btheta{}'\) \; d^2\btheta{}'
\label{eq-arriv}
\end{equation}
where $\kappa$ is the dimensionless density (also called convergence).
If we measure angles in arcsec, $\tau$ will have units of $\rm
arcsec^2$. To turn it into days we have to multiply by $g\,\Tzls$,
where $T$ is a time scale and $g$ is $H_0^{-1}$ in Gyr (see
Appendix~A).  In addition to the indicated redshift dependence,
$\Tzls$ has a weak cosmology dependence. By default {\em PixeLens\/}
assumes $\Omega_M=0.3,\Omega_\Lambda=0.7$, which gives
$T^{1115} = 3.34$~days, $T^{1608} = 10.28$~days, and $T^{1933} = 9.72$~days.

In our models, the mass is pixelated. Therefore we can simplify
Equation (\ref{eq-arriv}) to
\begin{equation}
\tau\(\btheta\) = \half|\btheta|^2 - \btheta\!\cdot\!\bbeta
- \sum_n\kappa_n Q_n\(\btheta\)
\label{eq-arrivpix}
\end{equation}
where $\kappa_n$ is the density on the $n$-th pixel and
$Q_n\(\btheta\)$ is the integral from (\ref{eq-arriv}) evaluated over
that pixel.  The explicit form of $Q_n\(\btheta\)$ is a messy but
straightforward mixture of logs and arctans and is given in
\cite{sw97}. The important thing about Equation (\ref{eq-arrivpix}) is
that it is linear in the unknowns, $\bbeta$ and $\kappa_n$.  We also
allow a constant external shear 
\begin{equation}
\gamma_1 (\theta_x^2-\theta_y^2) + 2\gamma_2 \theta_x\theta_y
\end{equation}
to be added to Equation (\ref{eq-arrivpix}); again, the shear
is linear in the unknowns $\gamma_1,\gamma_2$.

We now implement the following observational constraints.
\begin{enumerate}
\item An image observed at $\btheta\!_i$ implies
\begin{equation}
\bnabla\tau\(\btheta\!_i\)=0.
\end{equation}
In a multiple-image system all the images have the same unknown
$\bbeta$, so each multiple-image system introduces
$2(\<images>-\<sources>)$ constraints.
Thus a double provides 2 image constraints, a quad 6, while 1933+503
provides 14.
\item A time-delay measurement between images at $\btheta\!_i$ and
$\btheta\!_j$ implies
\begin{equation}
\tau\(\btheta\!_i\)-\tau\(\btheta\!_j\) = {\<obs delay> \over g\;\Tzls} .
\label{eq-obsdel}
\end{equation}
Here $g$ is another unknown and, fortunately for us, $g^{-1}$ appears
linearly.  If a time-delay is unmeasured, but we are sure about the
ordering, we replace (\ref{eq-obsdel}) by an inequality constaint
\begin{equation}
\tau\(\btheta\!_i\)-\tau\(\btheta\!_j\) \geq 0 .
\end{equation}
We have 3 equality constraints from time delays in 1608+656,
2 in 1115+080, and none in 1933+503.
\item We require the images to have the expected parities, and for the
image elongation when projected along the radial direction to be
between $1\over10$ and 10.  Although very modest, these requirements
do suppress a tendency for spurious images to be created in the
vicinity of a nearly-merging pairs of genuine images. Constraints of
this type are also linear \citep{asw98}.
\item We remarked after Equation (\ref{eq-obsdel}) that $g$ is an
unknown variable in the constraint equations.  Now, if we write the
constraint equations for several lenses, then each lens will introduce
a separate unknown variable for $g$, say $g^{1115}$, $g^{1608}$ and so
on.  But $g$ is universal, so we have additional constraint equations
of the type
\begin{equation}
g^{1115} = g^{1608} \, , \qquad g^{1608} = g^{1933} \, .
\label{hcoup}
\end{equation}
Though $g$ is still unknown, it now couples the constraint equations
for different lenses.  This coupling is the central astrophysical
contribution of this paper. In particular, it means that if one lens
rules out a certain range of $g$, {\em PixeLens\/} will not consider
models of any lens that require $g$ in that range.
\end{enumerate}

We also impose secondary constraints based on our general knowledge of
galaxy mass profiles.  
\begin{enumerate}
\item All the $\kappa_n \geq 0$,
\item If the galaxy does not appear very asymmetric, the mass profile
is required to have inversion symmetry about the lens center.
\item The density gradient anywhere must point within $\leq 45^\circ$
of the lens center.  The precise formulation of this constraint is
given in \cite{sw97}.
\item The $\kappa_n$ of any pixel must be $\leq 2\<average of
neighbors>$, except for the central pixel, which is allowed to be
arbitrarily dense.
\item The radial mass profile must be steeper than $|\btheta|^{-0.5}$.
The stellar dynamical evidence indicates that the total   
density distribution in the central regions of ellipticals is well
approximated by isothermal, $r^{-2}$ \citep{rix97,gerhard01},
while the study of the dynamics of the gas in the center of our Galaxy
show that total density scales as $r^{-1.75}$ \citep{binney91}. Guided by
these studies we choose $|\btheta|^{-0.5}$ as a conservative minimal 
steepness of the 2D density profiles in galaxies.

\end{enumerate}
In Bayesian terminology, these secondary constraints constitute a
prior.

\begin{figure}
\epsscale{0.6}
\plotone{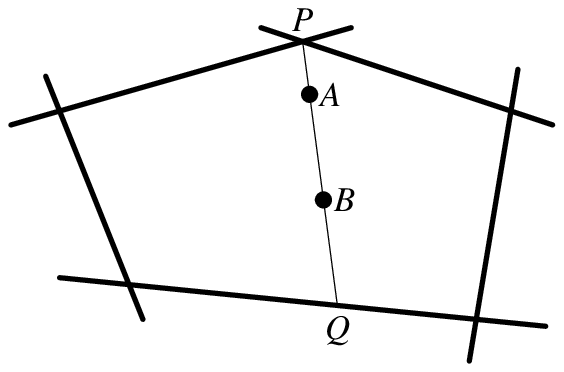}
\caption{Illustration of the Monte-Carlo procedure. The thick lines
represent inequality constraints, and any point in the pentagonal
region enclosed by these lines is an allowed solution.  Suppose we are
at solution point $A$.  We now do the following. (1)~Choose a random
vertex $P$ of the allowed region. (2)~Draw the line $PQ$ by drawing
$PA$ and extending it to end of the allowed region. (3)~Choose a
random point $B$ on $PQ$, and move to it.  The roles of $A$ and $B$ in
this sequence of operations is symmetric.  Hence the probability of
moving move $A\rightarrow B$ equals the probability of move
$B\rightarrow A$.}
\label{walk}
\end{figure}

The observational and prior constraints confine allowed lenses to a
polyhedron in the space of
$(\kappa_1,\kappa_2,\ldots,\kappa_N,\bbeta)$.  {\em PixeLens\/}
searches this polyhedron by a Monte-Carlo technique, briefly explained
in Figure \ref{walk}.  The key idea is to random-walk between models
such that probability of moving from model $A$ to model $B$ equals the
probability of moving from model $B$ to model $A$.  This is a simple
example of a Metropolis algorithm, or a Markov-chain Monte-Carlo
method.  Generalizations to non-uniform priors are also
possible---see, for example, \cite{pda}.

In our example, each model that {\em PixeLens\/} generates is a
three-lens model: there are mass maps, external-shear models where
required, and inferred source positions for each of 1115+080,
1608+656, and 1933+503.  In any one model the lenses are coupled by a
common value of $g$, but different models may have different values of
$g$.

As soon as the first model is ready, the user can examine it.  But the
users will not want to examine every model in an ensemble of 100.  So
{\em PixeLens\/} presents two kinds of results from the ensemble:
details of the average model of the ensemble-so-far and various
statistics derived from the ensemble. (By ensemble-average model we
mean a model with the densities in each pixel, the source positions,
and any external shear averaged over the ensemble.  The constraint
equations being linear in the quantities being averaged, they are
still satisfied by ensemble averages.)  Let us consider the two kinds
of results in turn.

\subsection{Results: the ensemble-average model}

\begin{figure}
\epsscale{0.38}
\plotone{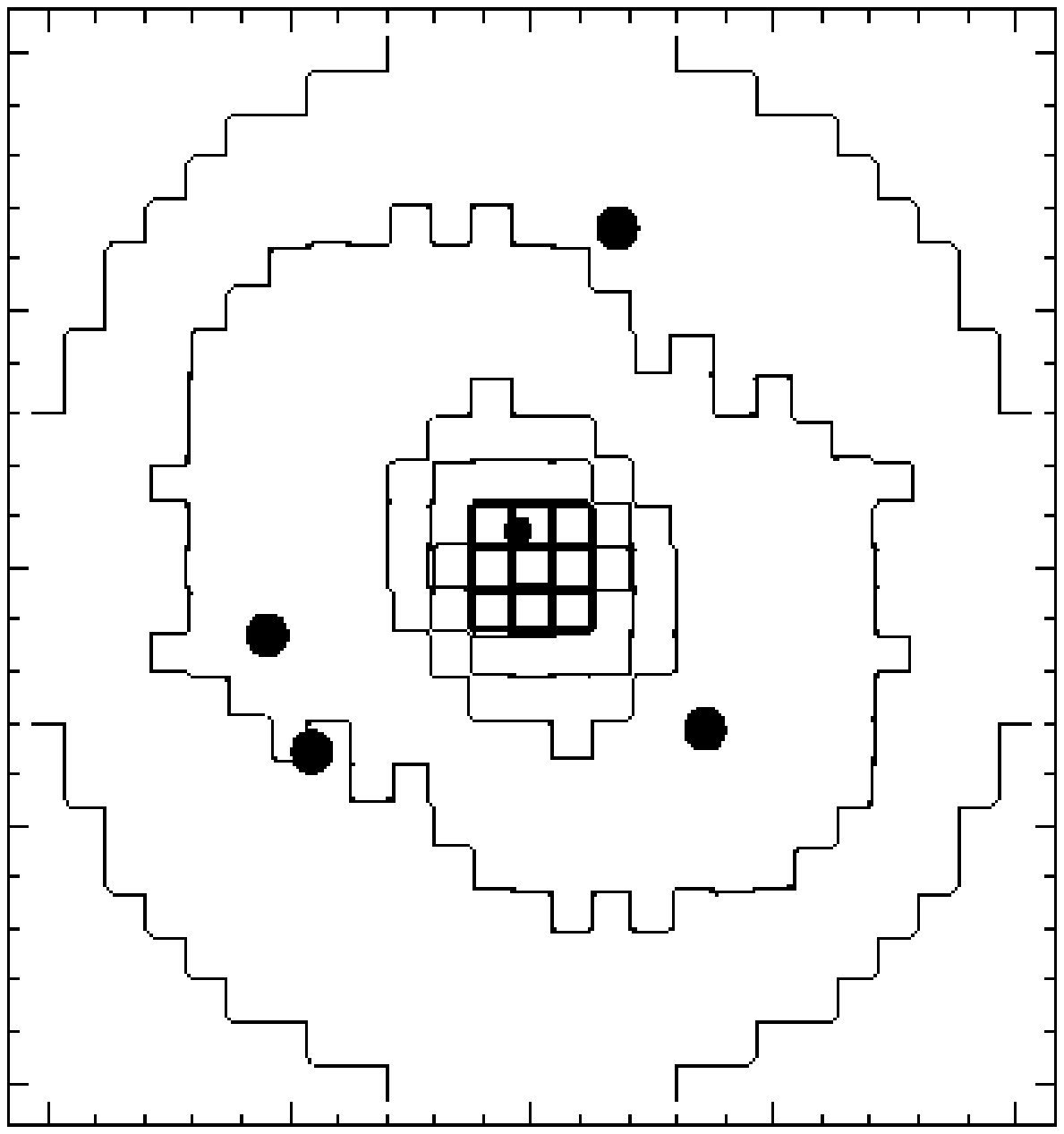}\par
\plotone{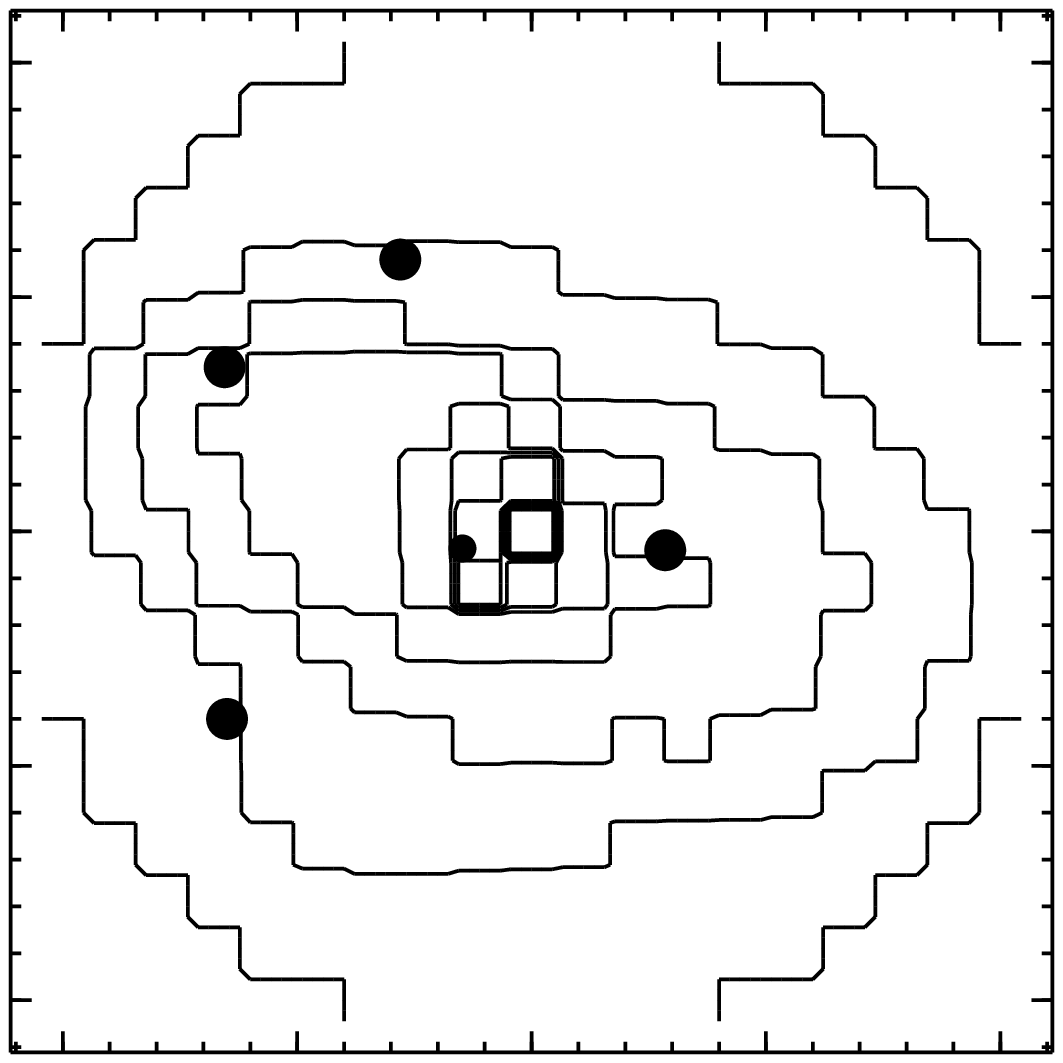}\par
\plotone{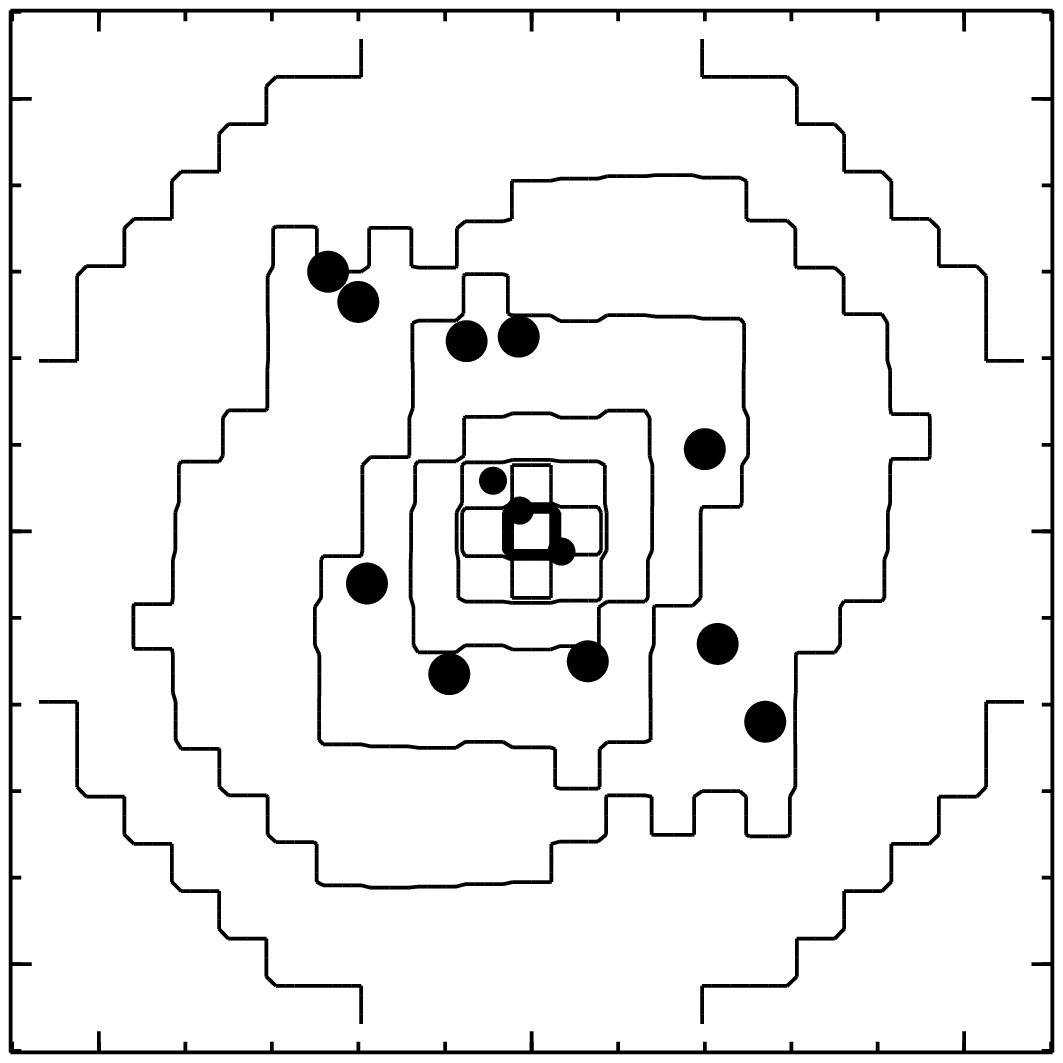}
\caption{Ensemble-average mass maps for the three lenses.  The
contours are in steps of $\kappa=\frac13,\frac23,\ldots$ The large
filled circles are the image positions. The small filled circles near
the centres are the inferred source positions.}
\label{mass}
\end{figure}

Figure \ref{mass} shows mass maps for the three lenses from the
ensemble-average model.  For 1115+080 and 1608+656, the mass maps are
very similar to those in \cite{ws00}.  The modeling method and input
data in the earlier paper were very similar---except for the coupling
(\ref{hcoup})---but the code was completely different.  We remark that
the elongation of the mass profiles in 1115+080 is towards the other
group galaxies, the elongations in 1608+656 and 1933+503 are what we
might expect from the light in those lenses.

\begin{figure}
\epsscale{0.38}
\plotone{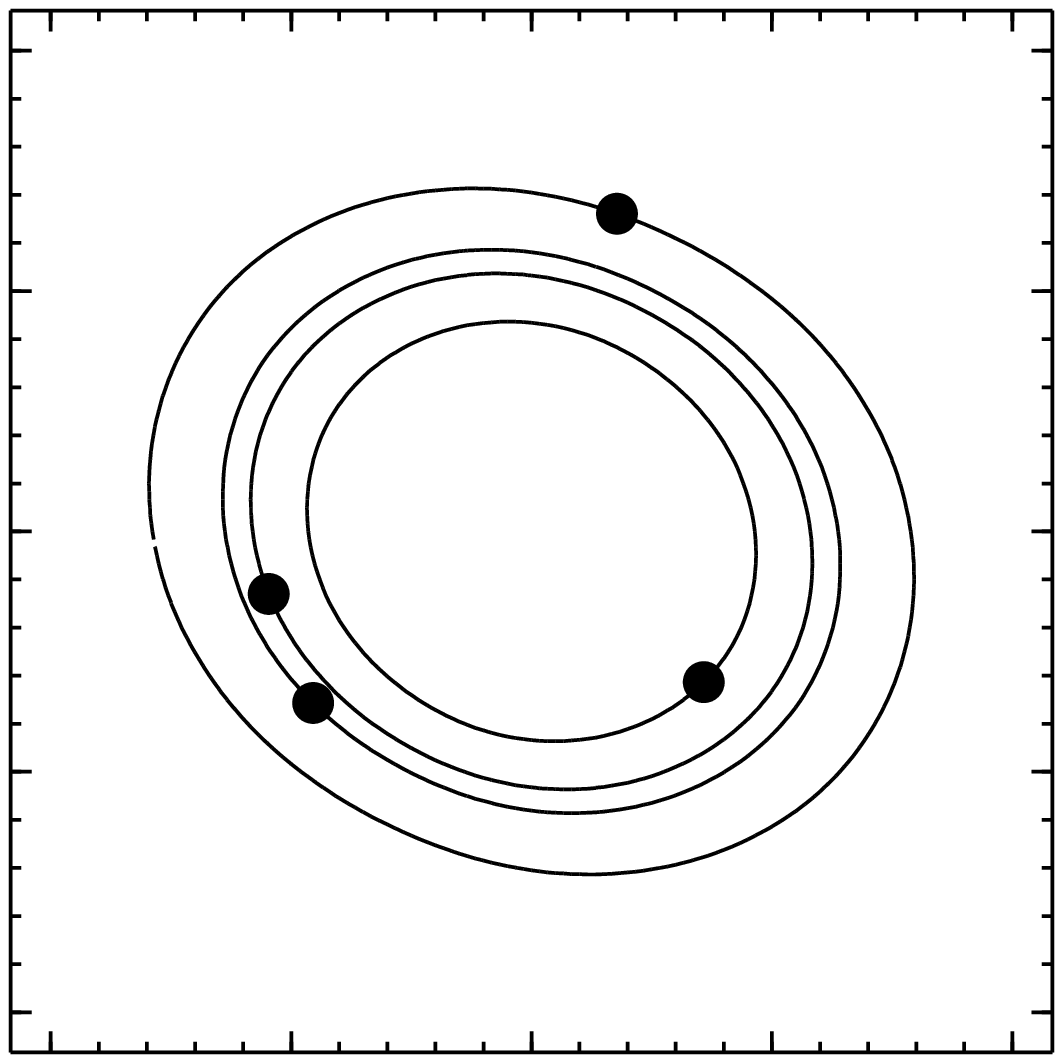}\par
\plotone{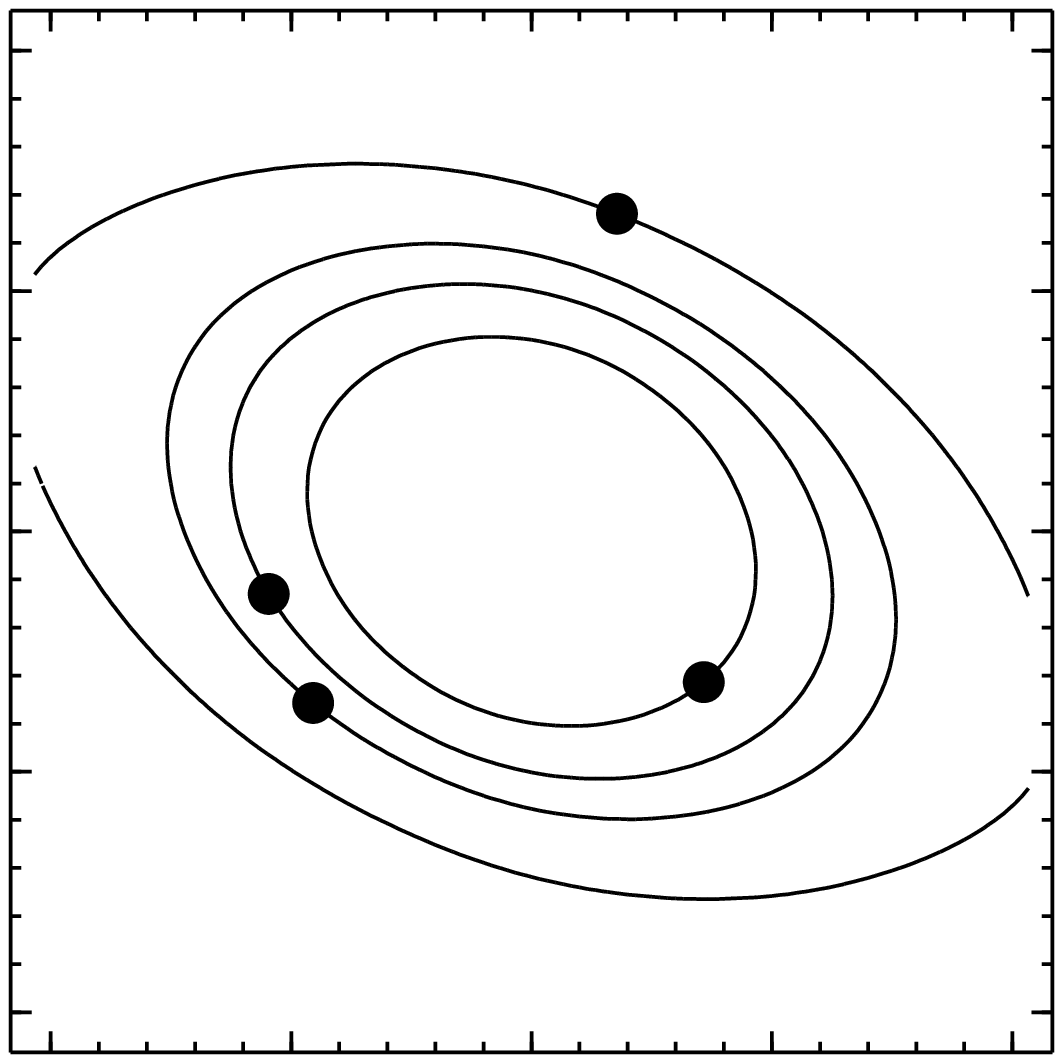}
\caption{Ensemble-average lens potential for 1115+080. Upper panel:
Contours of constant $\psi$, the contours shown being the ones passing
through the images. Lower panel: Like the upper panel, but with the
external shear exaggerated by a factor of 3.}
\label{poten}
\end{figure}

Figure \ref{poten} shows the lens potential in 1115+080.  (The default
is to pick the contours that pass through the images, but the user can
try different contour spacings interactively, by typing in the appropriate
values in the `plot parameters' boxes.) Notice how smooth the
potential is: nothing of the pixelation so evident in the mass maps
is visible here; it has been washed out by the double integral in
Equation (\ref{eq-arriv}).

Figure \ref{poten} also illustrates a convenient way of gauging the
importance of external shear in a model.  {\em PixeLens\/} allows the
user to plot the lens potential with the external shear exaggerated.
(Such exaggeration is purely for visualization, it has no effect on
the models.)  The user can adjust an exaggeration parameter: the
default is 1, while 0 means omit the external shear.  Figure
\ref{poten} shows that a 3-fold exaggeration makes a drastic
difference to the potential, indicating that the external shear in
this lens is large but not dominant.

\begin{figure}
\epsscale{0.38}
\plotone{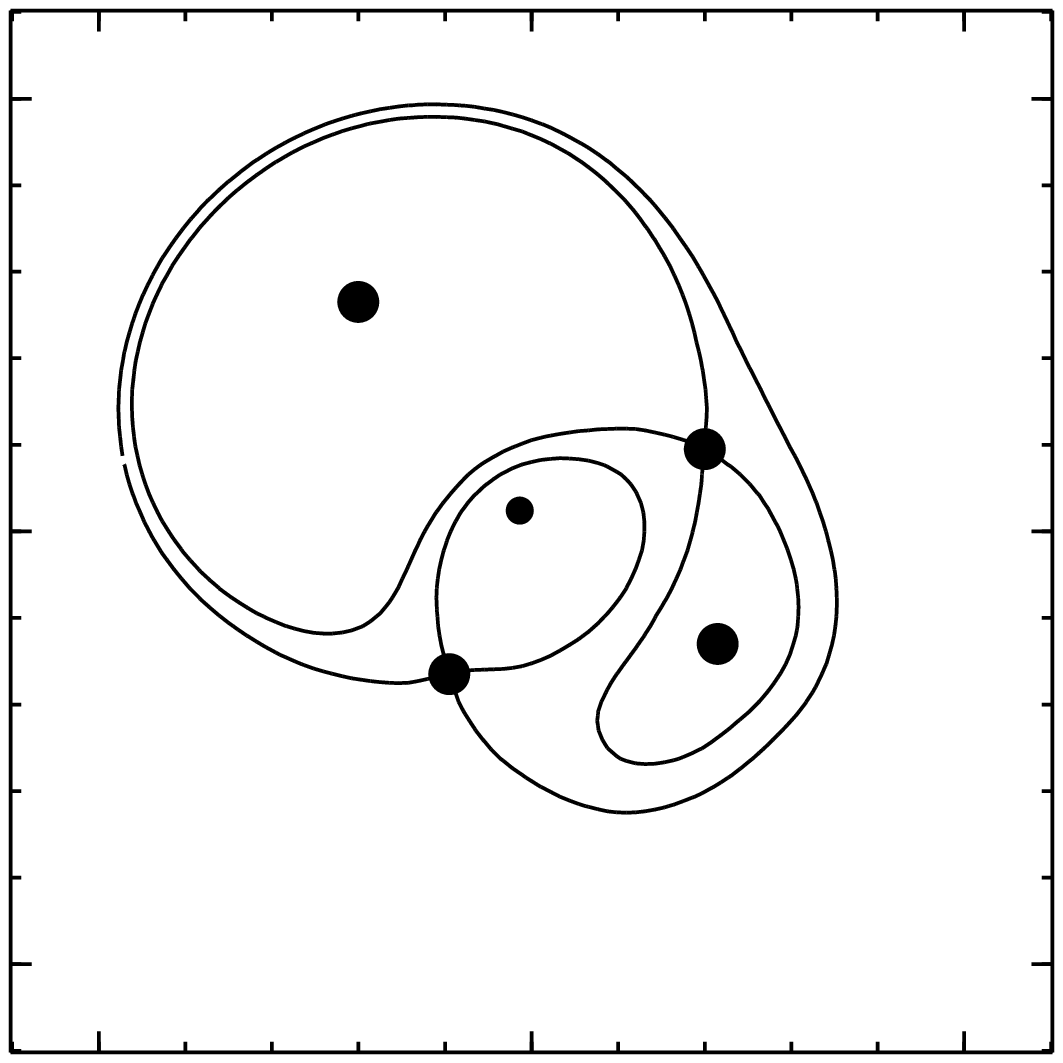}\par
\plotone{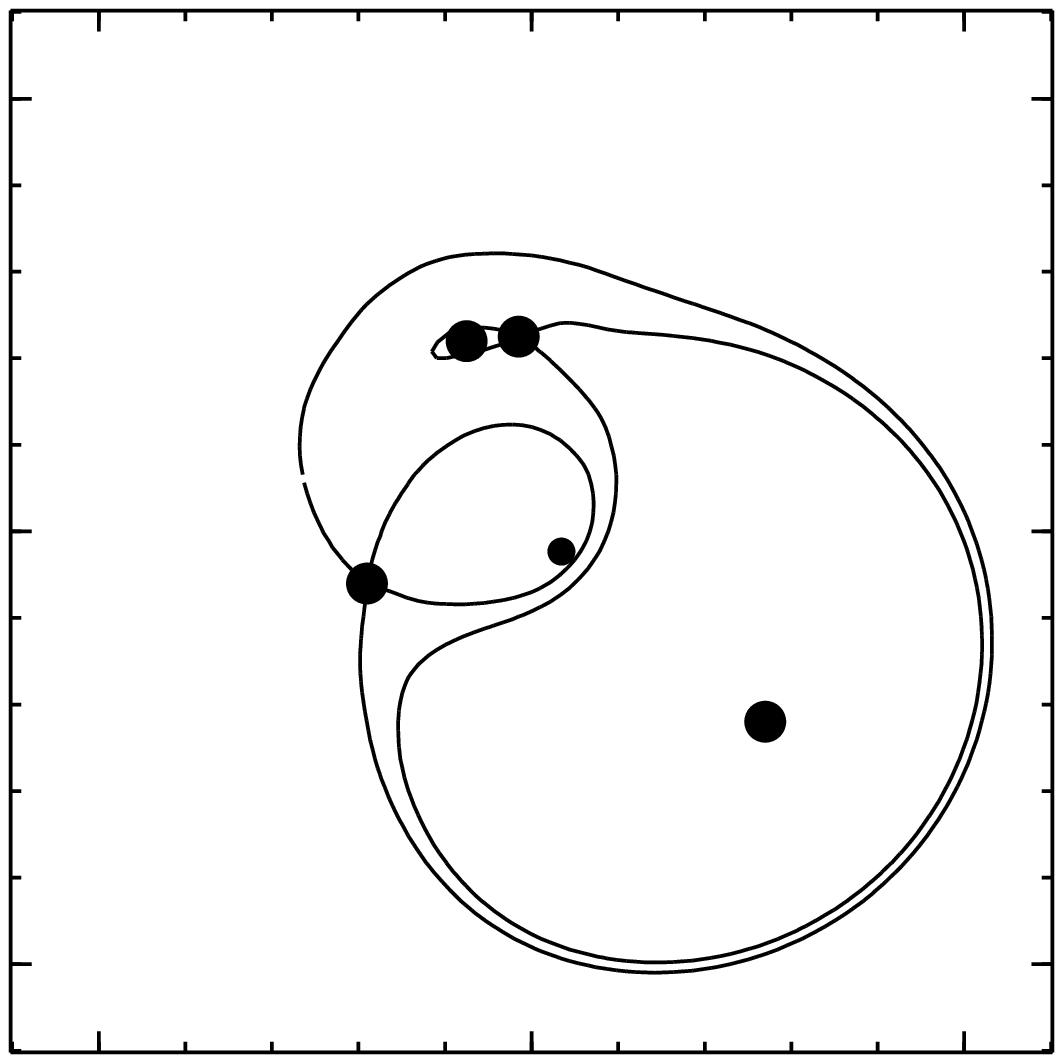}\par
\plotone{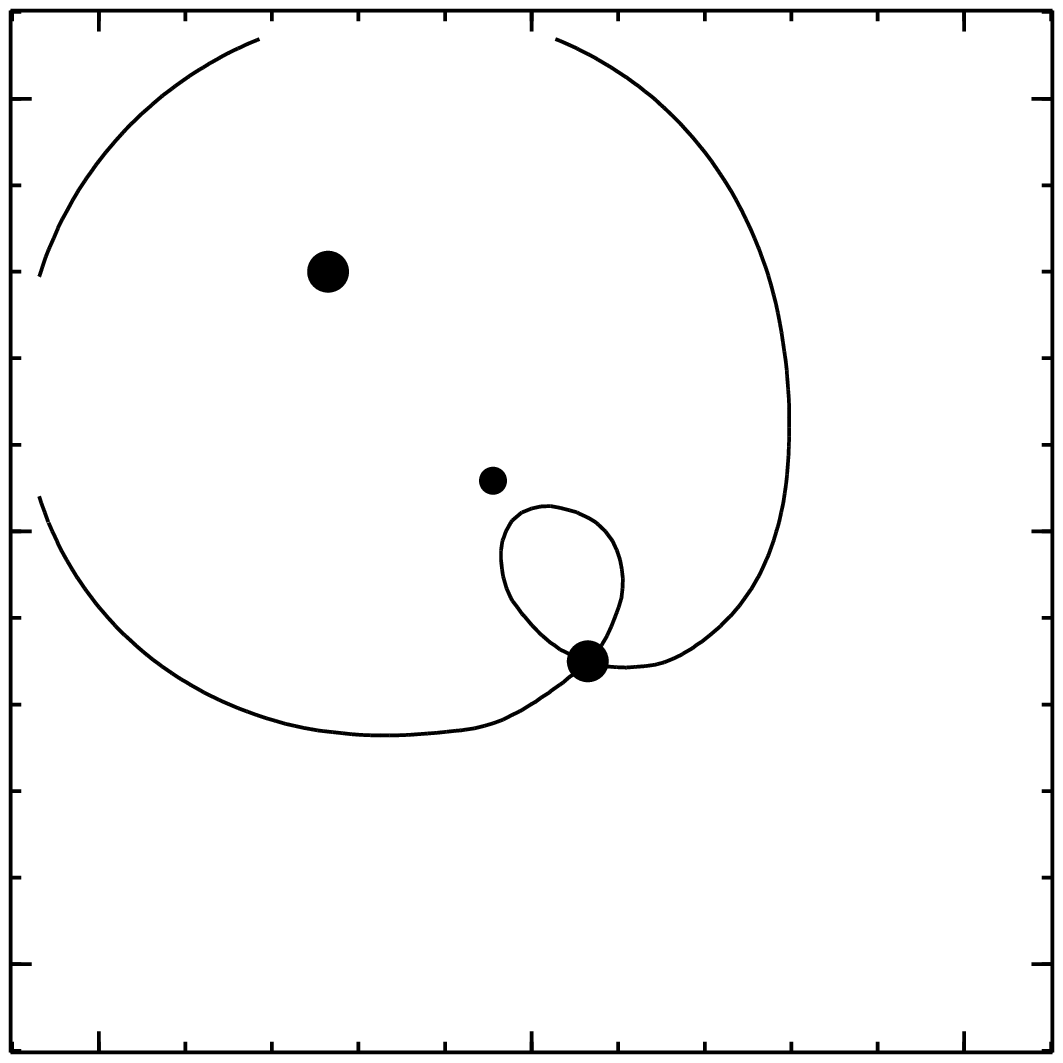}
\caption{Ensemble-average arrival times for the three image systems in
1933+503: core quad (upper), lobe quad (middle), and lobe double
(lower). Only the saddle-point contours are shown.}
\label{arrivb}
\end{figure}

Figure \ref{arrivb} show saddle-point contours in the arrival-time
surfaces for the two quads and a double of 1933+503.  Saddle-point
contours are the default, but as with the potential, the user can try
different contour spacings interactively. Examining the arrival-time
contours is a very good way of checking that the model is sensible:
small errors in the input or the modeling code nearly always lead to
noticeable spurious features in the arrival-time surface, such as
contorted contours or extra images.

\begin{figure}
\epsscale{0.6} 
\plotone{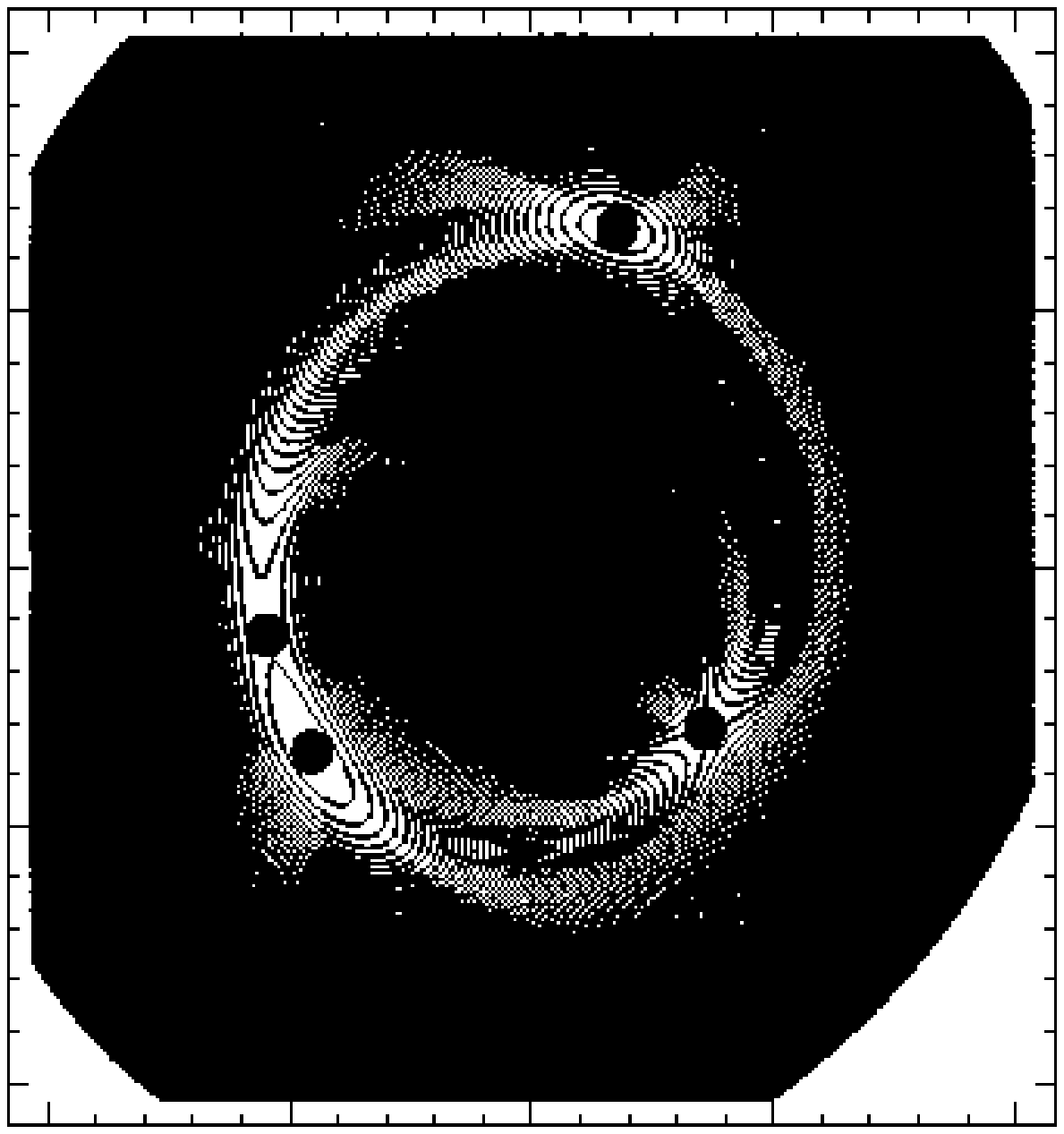}
\caption{Arrival-time surface for 1115+080, with contours spaced by
$0.005\rm\,arcsec^2$.  This represents a simple model of the Einstein
ring formed by the host galaxy.}
\label{arrivc}
\end{figure}

Arrival-time contours also have another, somewhat surprising, use.
Figure \ref{arrivc} shows the result if the user specifies a
contour-spacing of 0.004.  (The units here are $\rm arcsec^2$. We can
convert to physical time units, as explained after Equation
\ref{eq-arriv}, and the result is about $5\rm\,hr$, but let us
continue with $\rm arcsec^2$.)  A plot of the arrival-time surface
with closely-spaced contours tends to resemble an Einstein ring.  In
fact, it approximates---see \cite{sw01} for details---the image of a
source with a conical light profile of radius
\begin{equation}
      {\<contour spacing>\over\<thickness of contour lines>} .
\end{equation}
Now, the line thickness in all the plots in this paper is $1\over300$
of the plot width.  For Figure \ref{arrivc}, that amounts to
$\simeq0.015''$. Hence Figure \ref{arrivc} predicts the image due to a
conical profile of radius $\simeq0.33''$ centered on the QSO.  In fact it
resembles the ring imaged by \cite{impey98}, though there are
differences of detail.  Thus, we can make quantitative inferences
about the host galaxy simply by inspecting images and doing some
mental arithmetic.

\subsection{The model ensemble: stats}\label{stats}

Out of the full ensemble of models, it is possible to extract many
kinds of statistics. {\em PixeLens\/} implements four kinds, all
involving $H_0^{-1}$.

\begin{figure}
\epsscale{0.6} 
\plotone{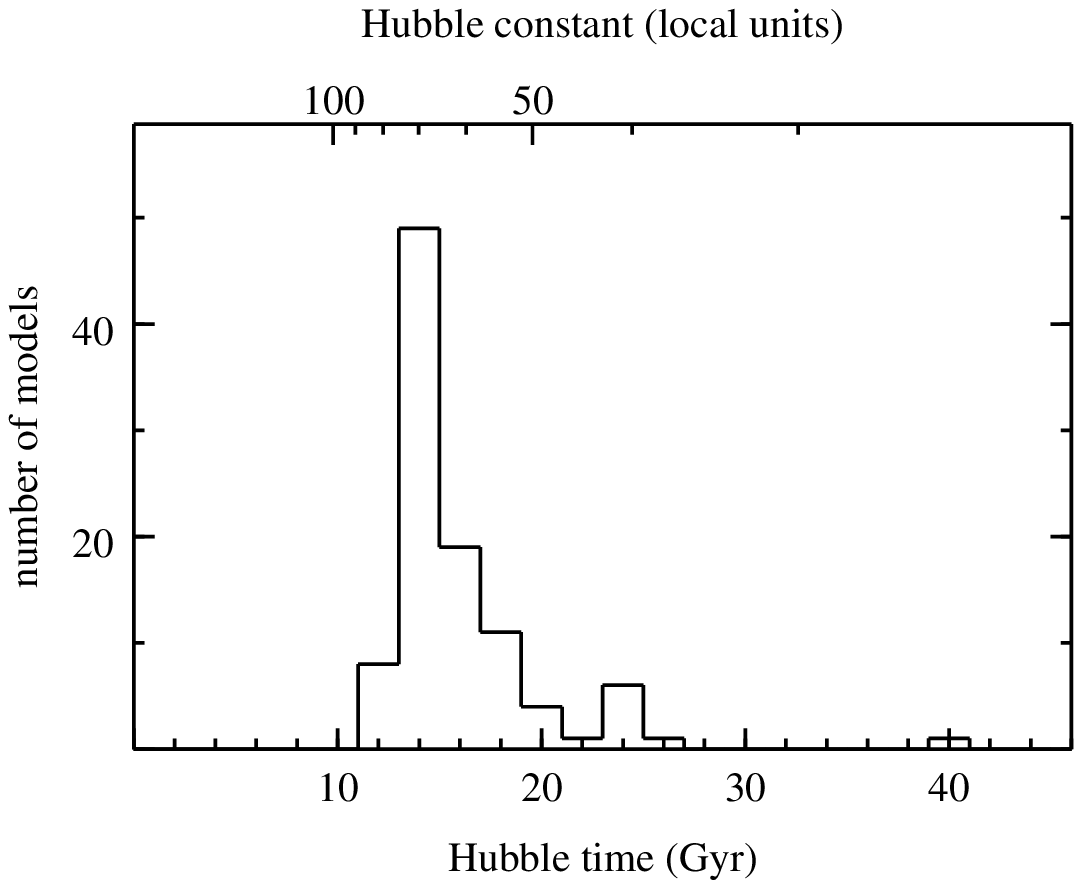}
\caption{Histogram of the Hubble time from 100 three-lens models.}
\label{nh}
\end{figure}

Figure \ref{nh} shows a histogram of $g$ values from the ensemble, and
has a straightforward interpretation as the posterior distribution of
$H_0^{-1}$.  This figure is analogous to Figure~17 in \cite{ws00}, but
better.  In the earlier work we had separate model ensembles for
1115+080 and 1608+656 and multiplied their $h$-histograms bin-by-bin.
That procedure had the problems of (i)~a reduced number of models in
the common region, and (ii)~increased shot noise from bin-by-bin
multiplication.  In {\em PixeLens,} because of the constraint
(\ref{hcoup}), there is only one histogram of $g$, which eliminates
both problems.  This histogram (or rather, the unbinned values it
represents) gives
$$ \matrix{
  {H_0}^{-1}=14.6_{-1.2}^{+3.8} \rm\ Gyr
& (H_0=67_{-14}^{\,+\;6} \rm\ local\ units)
& \hbox{at 68\% confidence} \cr
  {H_0}^{-1}=14.6_{-1.7}^{+9.4} \rm\ Gyr
& (H_0=67_{-26}^{\,+\;9} \rm\ local\ units)
& \hbox{at 90\% confidence.} \cr
}$$ 

The median $H_0$ is somewhat higher than in \cite{ws00}, but this is
expected because that paper used an Einstein-de-Sitter cosmology,
which shortens $\Tzls$ by $\sim5\%$ for 1115+080 and by $\sim10\%$ for
1608+656.  The uncertainties are similar.

\begin{figure}
\epsscale{0.47}
\plotone{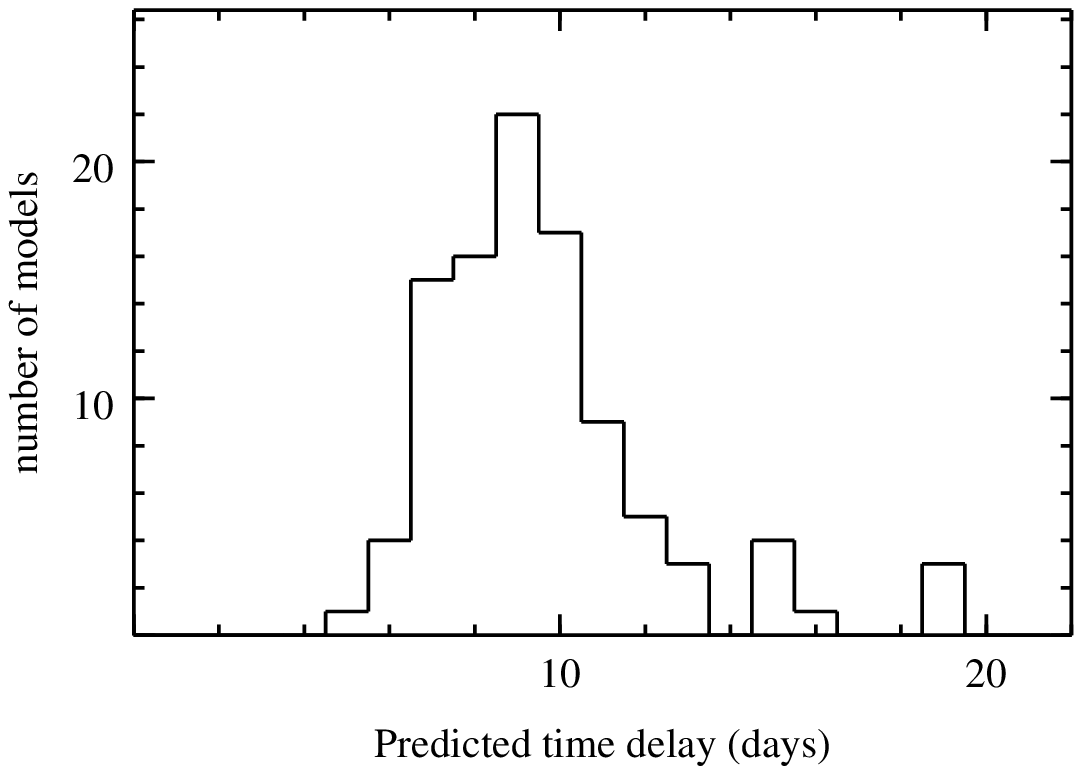}\par
\plotone{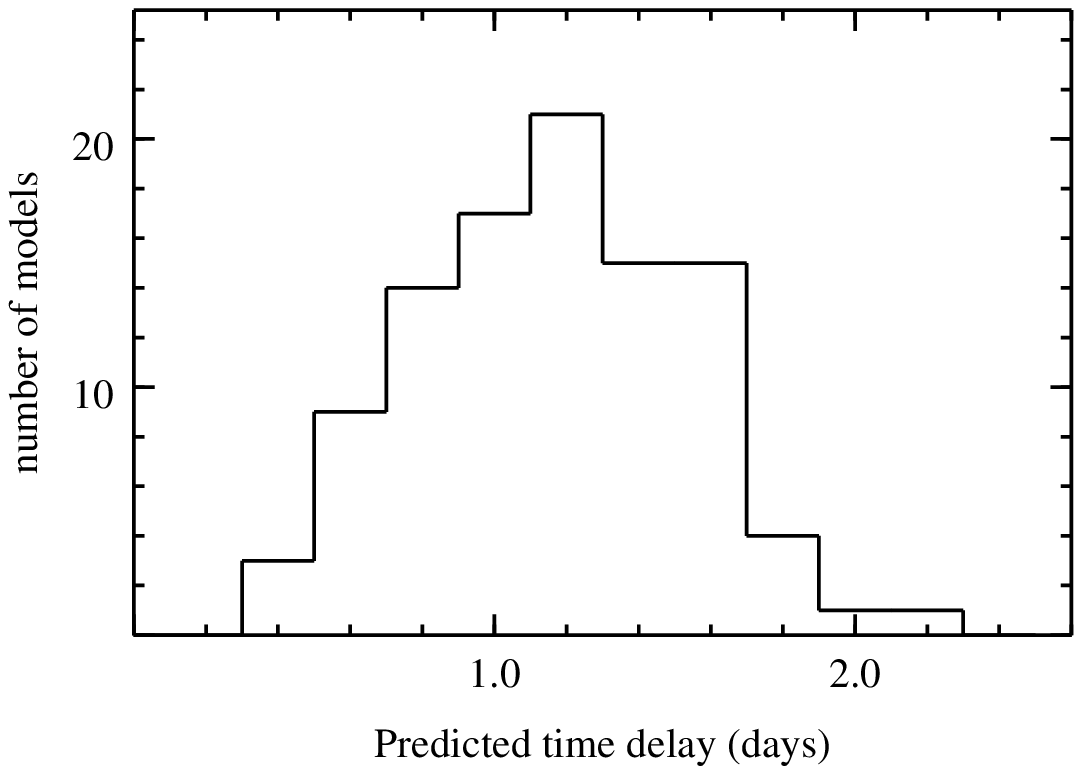}\par
\plotone{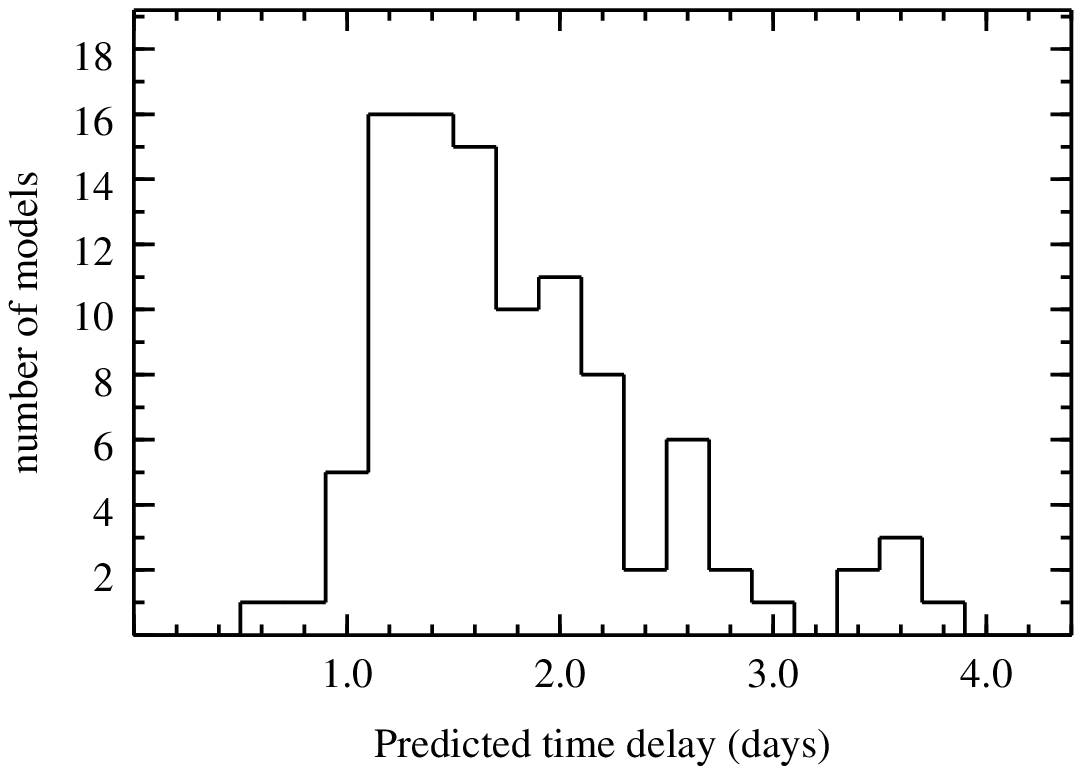}
\caption{Time-delay predictions for the core quad in 1933+503: between
images 1 and 2 (upper panel), between 2 and 3 (middle), and between 3
and 4 (lower). See \citet{sw03} for an explanation of how to number
images based on arrival times.}
\label{del}
\end{figure}

Figure \ref{del} shows the predicted time-delays for the core-quad in
1933+503.

\begin{figure}
\epsscale{0.47}
\plotone{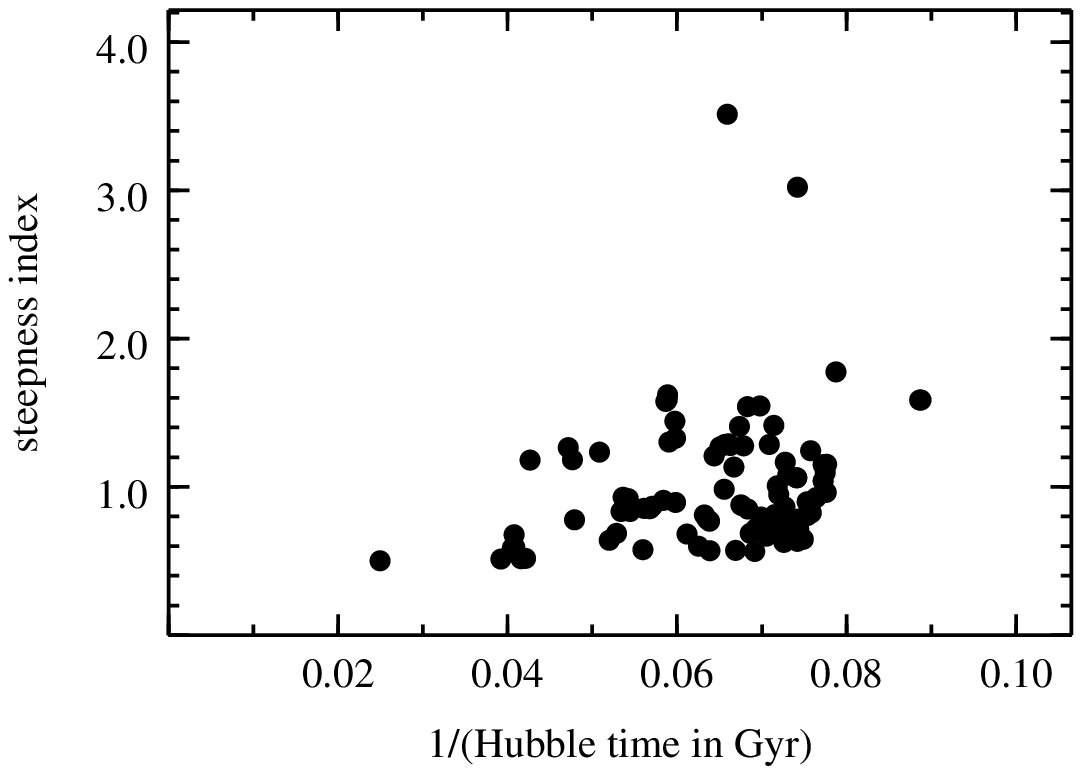}\par
\plotone{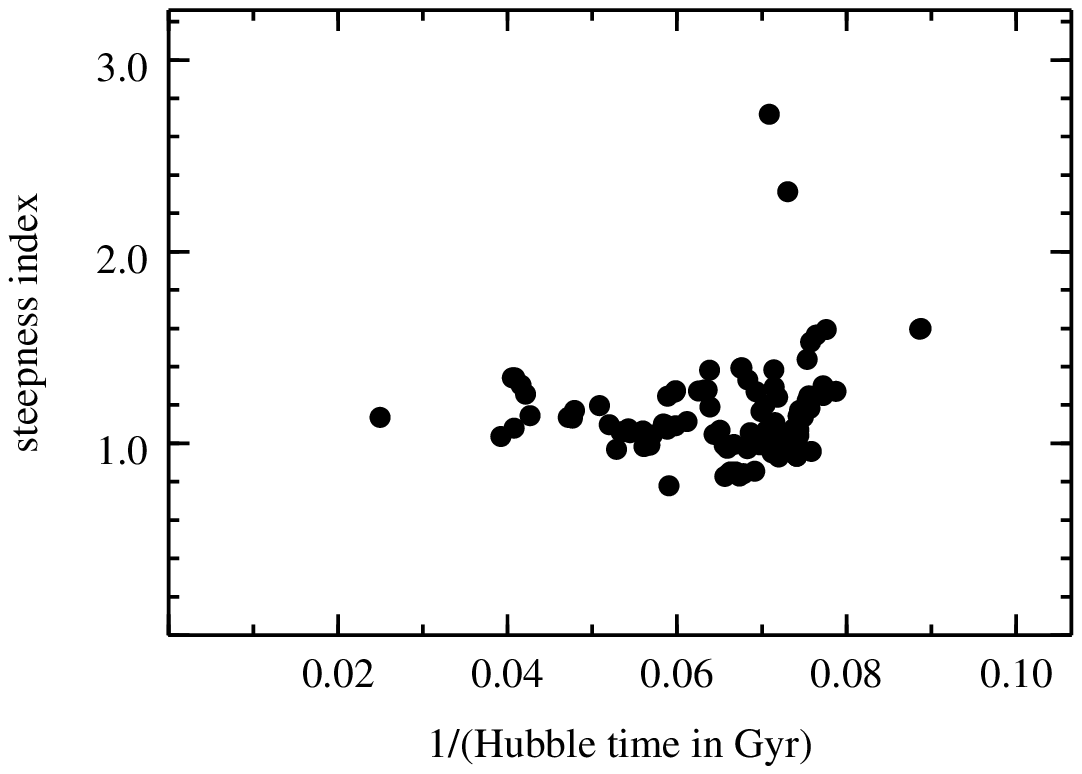}\par
\plotone{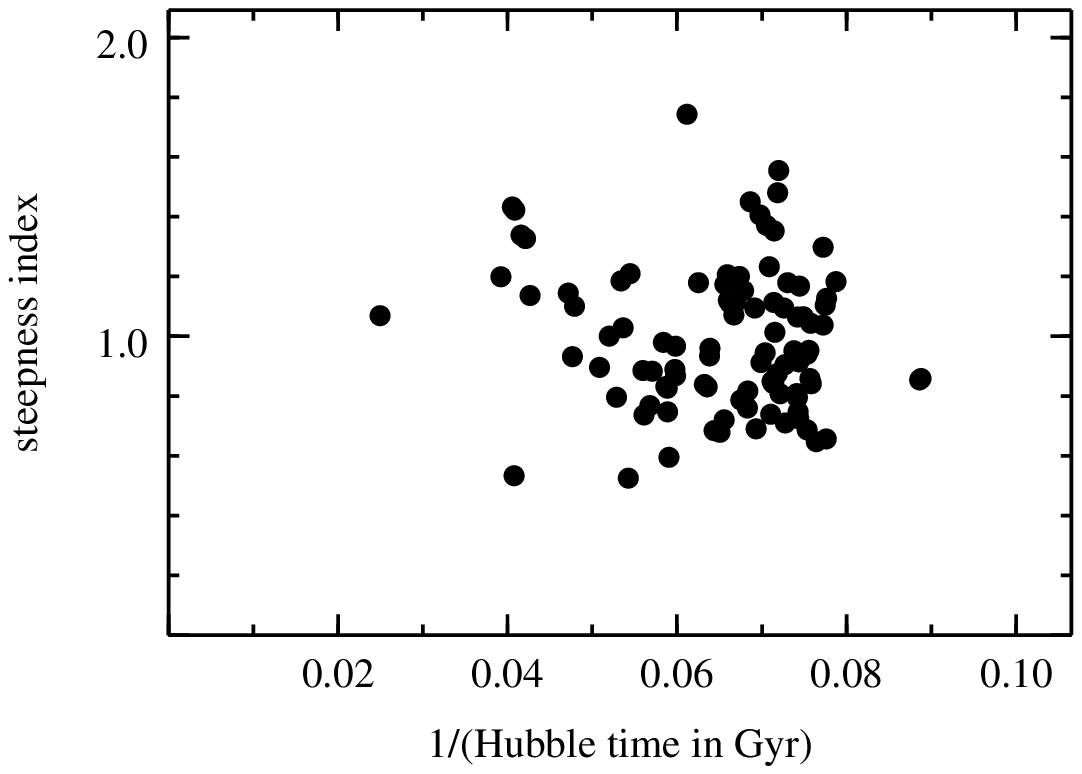}
\caption{Steepness index $\alpha$ against $g$, for 1115+080 (upper
panel), 1608+656 (middle), and 1933+503 (lower).  The upper and middle
panels are analogous to Figures 13a and 16a respectively in
\cite{ws00}, except for an obvious change of sign.}
\label{kalf}
\end{figure}

Figure \ref{kalf} shows the radial index $\alpha$ from a radial fit of
$|\btheta|^{-\alpha}$ against $g^{-1}$.  Most lens systems when
modeled independently show a correlation between $\alpha$ and $g^{-1}$
(or $h$), in the sense that steeper density profiles result in higher
values of $h$ (e.g., Figure 13a and 16a of Williams \& Saha 2000).
For the three coupled lenses shown here, 1115+080 shows the
correlation weakly and the others hardly at all.

\begin{figure}
\epsscale{0.47}
\plotone{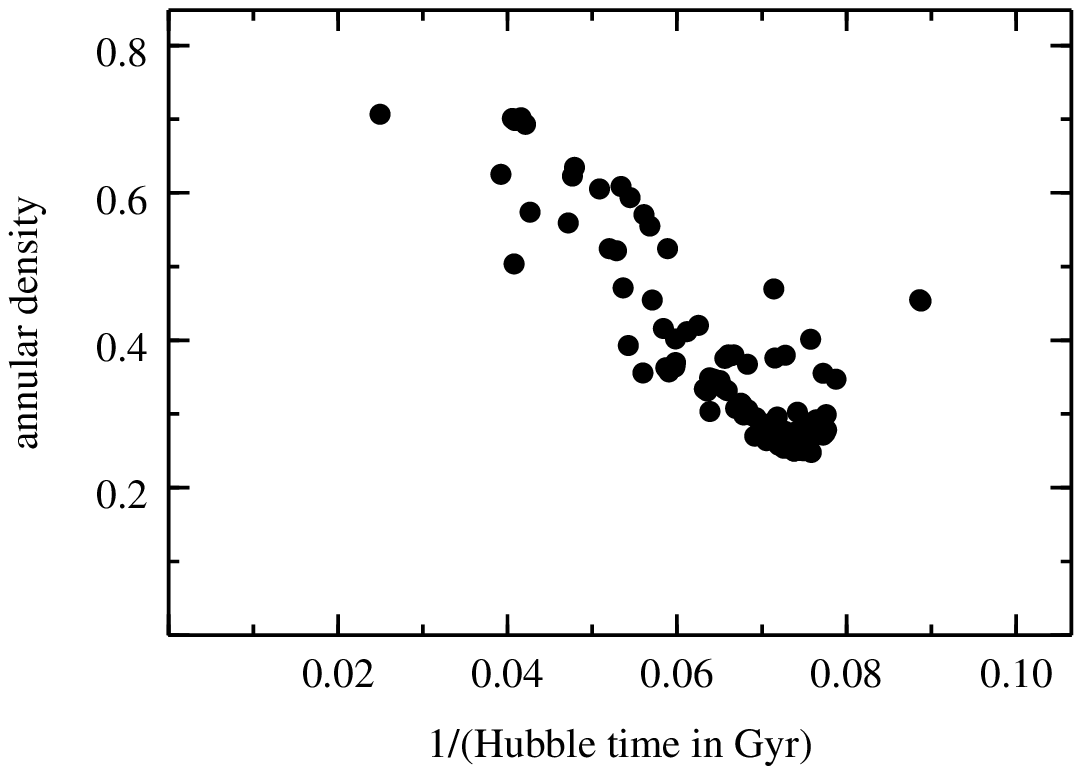}\par
\plotone{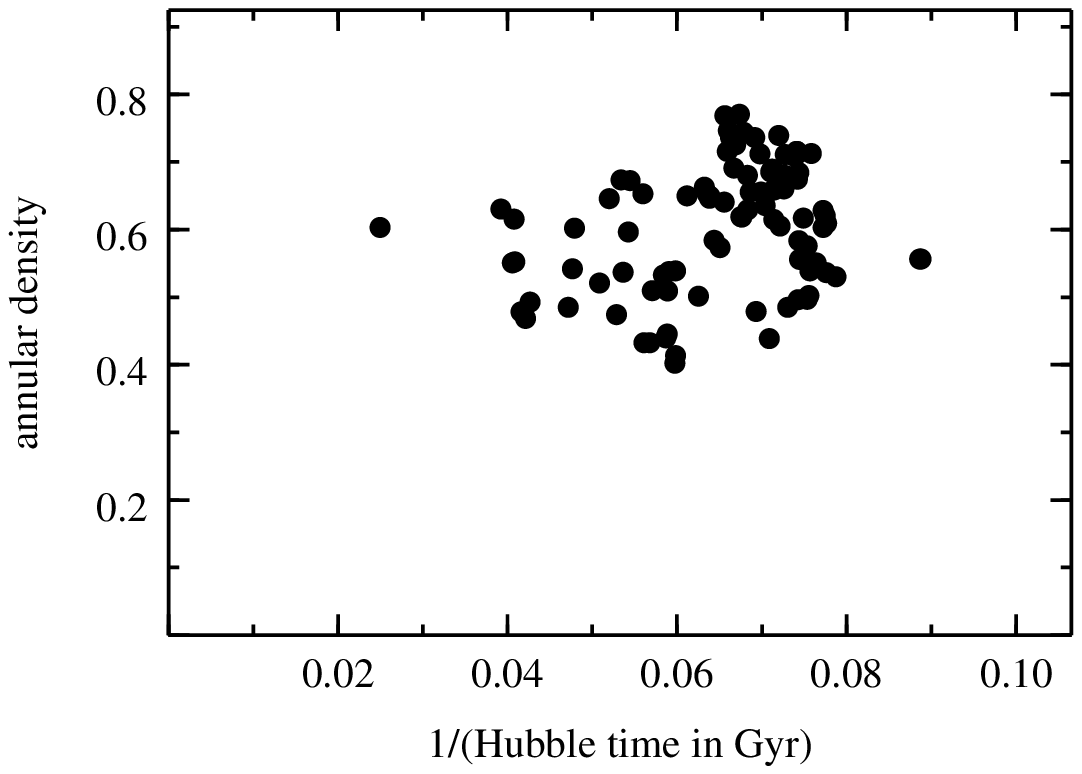}\par
\plotone{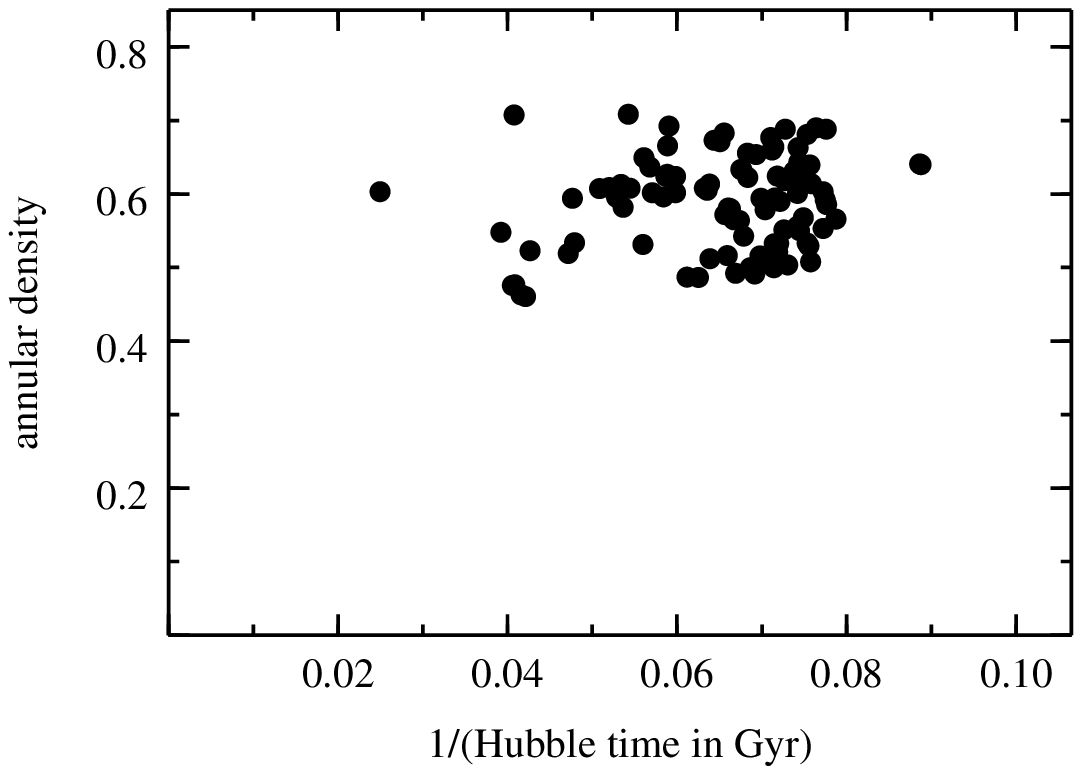}
\caption{Annular density $\kann$ against $g^{-1}$, again for 
1115+080 (upper panel), 1608+656 (middle), and 1933+503 (lower).}
\label{kann}
\end{figure}

Figure \ref{kann} shows the mean $\kappa$ in the annulus between the
innermost and outermost images (say $\kann$) against $g^{-1}$.
Again, when modeled independently, most lens systems show an anti-correlation
between $\kann$ and $g^{-1}$, whereas two of the three panels of 
Figure \ref{kann} show no trends at all. This can be expected, because
coupling different time-delay systems weakens correlations:
coupling reduces the range of $g$ explored, and over a
reduced range of $g$ any correlation involving $g$ will appear weaker.
For an extreme case, suppose we artificially fix the value of $g$,
which we can do by adding (say)
\begintype
g 14
\endtype
at the end of the input.  Then Figures~\ref{kalf} and \ref{kann} will
consist of vertical clusters of point at $g=14$.

However, the above argument does not fully account for the lack of
trends in the middle and lower panels of Figure \ref{kann}; other
causes are partially responsible. 
\cite{kochanek02} derives a linear relation between time delay,
$\kann$, and $h$, as a leading order approximation from lensing
theory. If one of these three (sets of) variables is fixed, say if 
the time delays are fixed by observations, then a well defined
trend would be expected to relate the other two ($\kann$ and $h$).
This is demonstrated by 1115+080. 1933+503, on the other hand, is not 
expected to exhibit such a trend because no time delays exist
for this system. Finally, 1608+656 shows no trend probably because it
is a distinctly asymmetric lens.  

\subsection{Artifacts}

To finish this worked example, we remark on some artifacts evident in
pixelated models.

The most obvious spurious feature is the pixel-scale substructure.  In
principle, one could replace mass pixels with basis functions, thereby
preserving the linearity of the constraint equations and leaving the
main algorithms unchanged, but making the mass maps smooth.  That
would be a useful development for microlensing applications, where
local density and shear are crucial.  But for studying image positions
and time delays, basis functions are unlikely to make any difference,
because the pixel scale structure is completely suppressed in the lens
potential and arrival time---see Figure~\ref{poten}.

Pixelation has a serious side-effect, however, if the pixels are made
too large.  We have found from experimenting with different pixel
sizes that if the pixels are too large and the mass-distribution is
poorly resolved, fewer steep mass-models are allowed.  As a result,
the prior is shifted to favor lower values of $g^{-1}$, and hence the
Hubble-time histograms also shift towards larger $H_0^{-1}$.  The
effect is most noticeable in the $\alpha$-$g^{-1}$ correlation.  The
pixel-sizes used for this paper are chosen small enough not to cause
this particular problem.

A second artefact concerns the central image.  Now the central maximum
itself is obvious from arrival time contour maps.  What is further
often evident is that the central maximum is slightly offset from the
lens center and not as steep as expected from the practical
unobservability.  This happens because our mass maps lack a central
density cusp, instead distributing the mass over the central pixel.
In principle, we could incorporate central density cusps into {\it
PixeLens.}  But it would make no difference to the observed images
because, as is well known, a circular mass disk has the same lensing
effect outside itself as a point mass.

A third (and rare) artefact is spurious extra images.  The
density-gradient constraint, we have found, is generally enough to
suppress extra images, but it is possible that a few rogue models with
spurious extra images are present in the ensembles.  In principle we
could reject those models by examining every single arrival-time
surface by eye, but we have not implemented that.

\section{A complementary worked example: four lenses}

In our second example we will be more brief.

We reconstruct four time-delay doubles.
\begin{itemize}
\item 1520+530 was discovered by \cite{chavushyan97} and \cite{burud02b}
measured its time-delay.
\item 1600+434 was discovered by \cite{jackson95} and \cite{burud00}
measured its time delay.
\item 1830-211 was discovered by \cite{rao88} and \cite{lovell98}
measured its time delay.
\item 2149-275 was discovered by \cite{wisotzki96} and \cite{burud02a}
measured its time delay.
\end{itemize}

\subsection{The data input}

We have already explained the input format, and without further ado
here are the data we input.
\begintype
object 1520                                   \hfil\break
symm pixrad 8 redshifts 0.71 1.855 shear 90   \hfil\break
double 1.141 0.395 -0.288 -0.257 130          \hfil\break
object 1600                                   \hfil\break
symm pixrad 8 redshifts 0.42 1.59             \hfil\break
double 0.610 0.814 -0.110 -0.369 51           \hfil\break
object 1830                                   \hfil\break
symm pixrad 8 redshifts 0.89 2.51             \hfil\break
double -0.50 0.46 0.15 -0.26 26               \hfil\break
object 2149                                   \hfil\break
symm pixrad 8 redshifts 0.490  2.03           \hfil\break
double 0.736 -1.161 -0.173 0.284 103 
\endtype
We can omit the number of models since 100 is the default.

In the case of PKS 1830-211, the position of the lensing galaxy at $z=0.89$ 
remains uncertain. Various scenarios have been proposed in the literature 
\citep{lehar00,winn02,courbin02,kdb03}; we use the astrometry from \cite{lehar00}.

\subsection{Results}

Since readers will have already seen the various kinds of results
possible, we now give only a small selection.

\begin{figure}
\epsscale{0.38}
\plotone{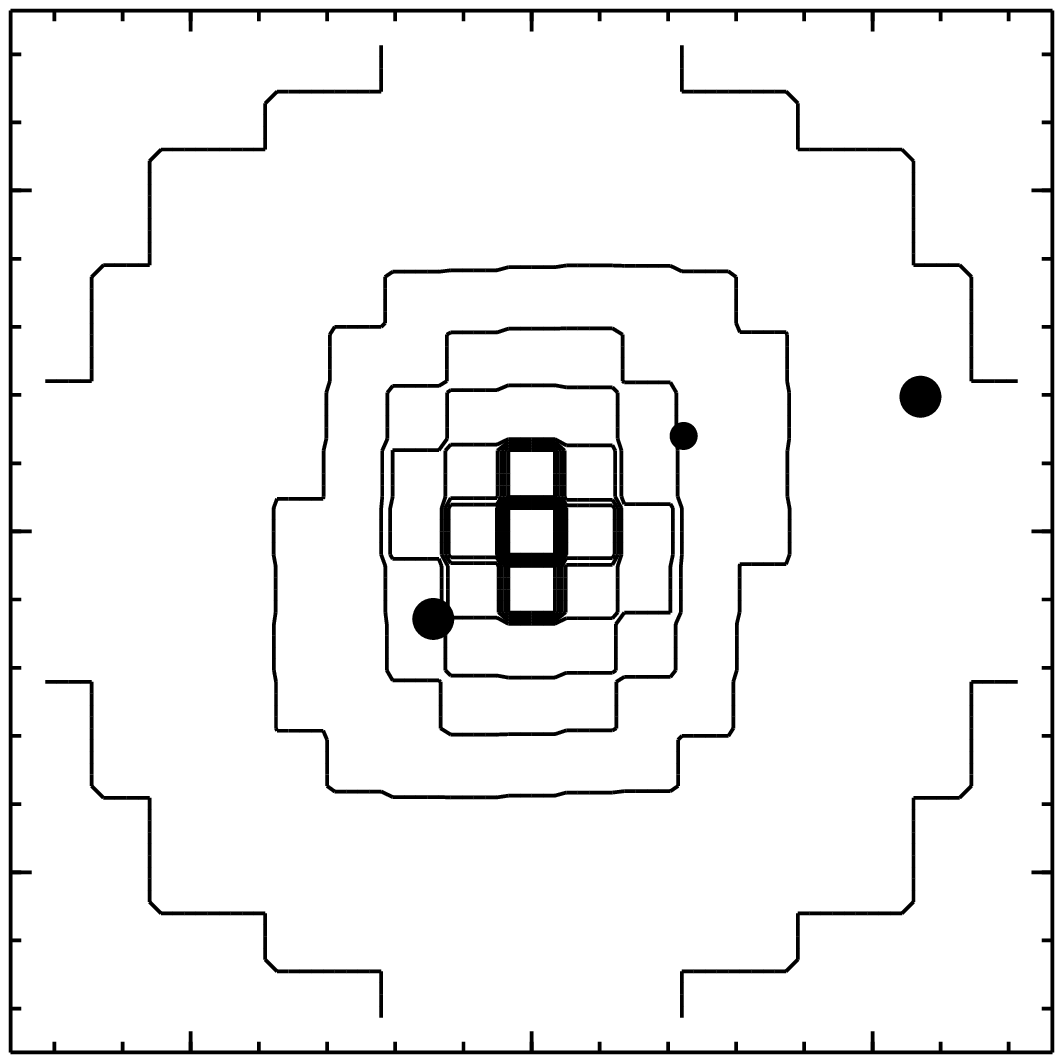}\plotone{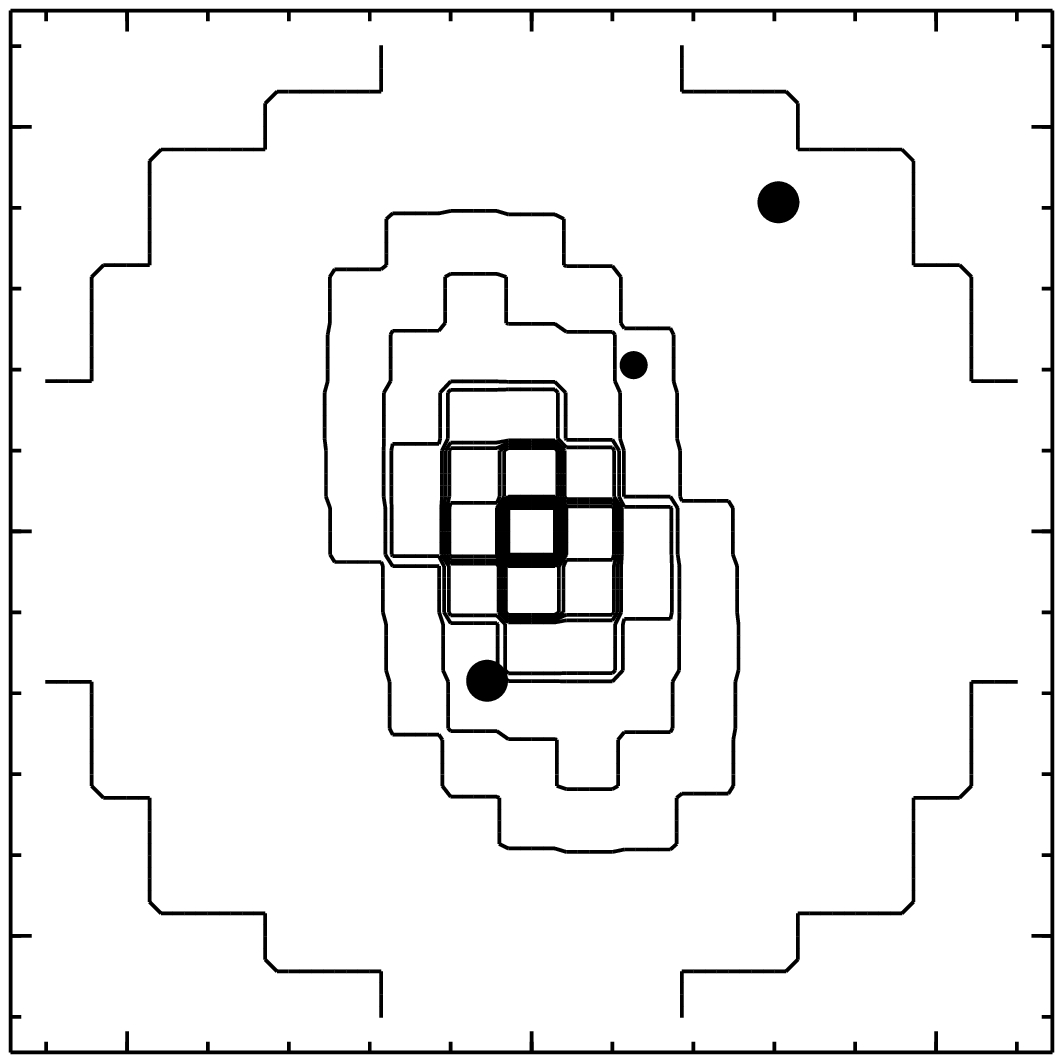}\par
\plotone{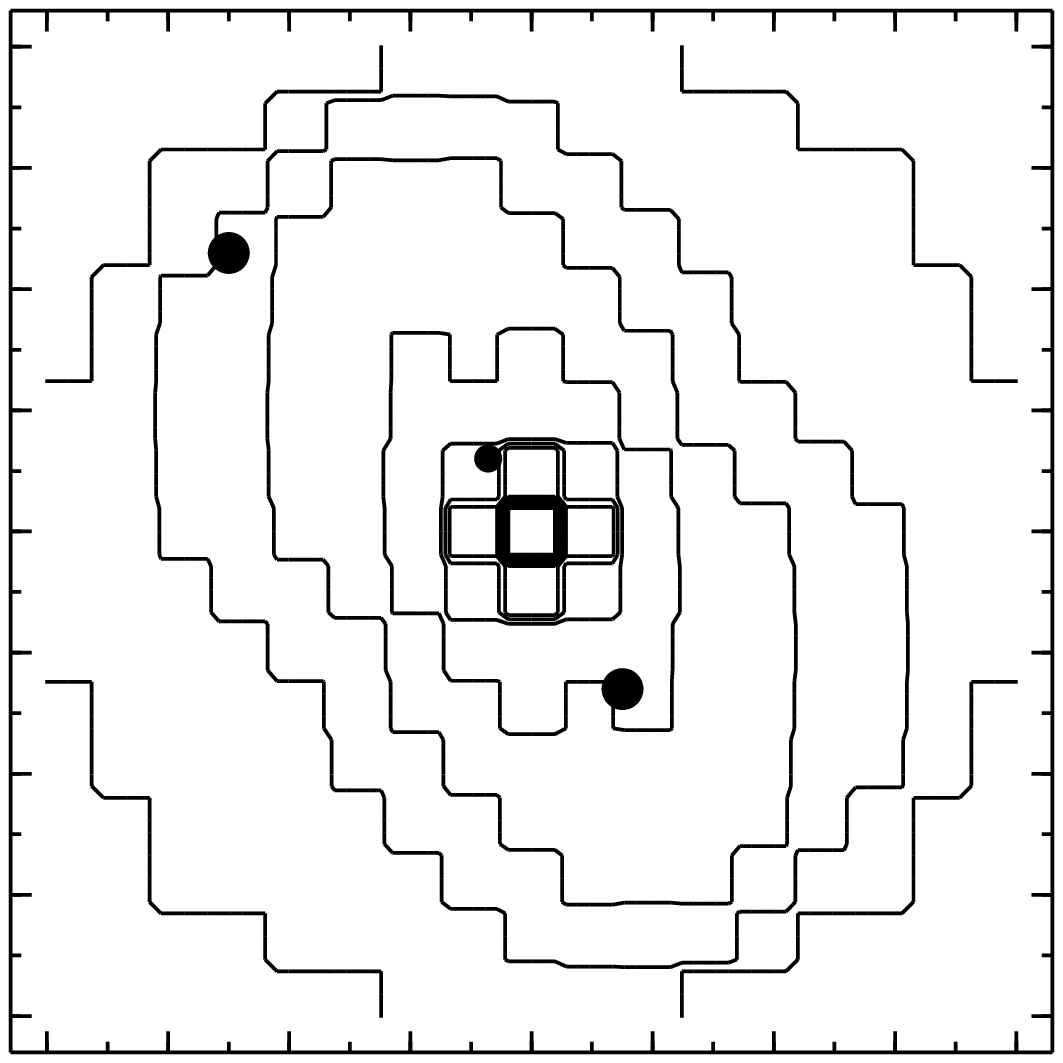}\plotone{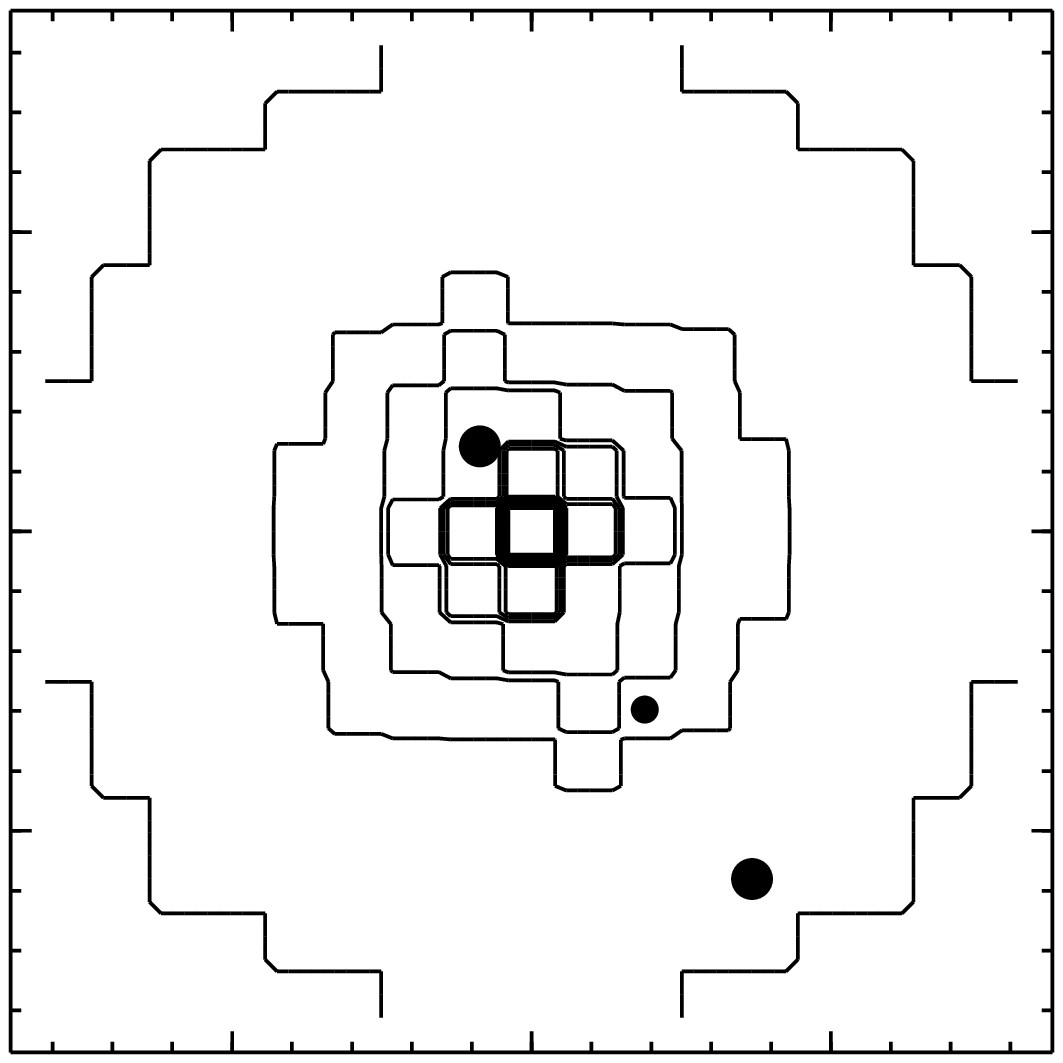}
\caption{Mass maps as in Figure \ref{mass}, but for
1520+530 (upper left), 1600+434 (upper right), 1830-211 (lower left),
and 2149-275 (lower right).}
\label{mass2}
\end{figure}

Figure \ref{mass2} shows the ensemble-average mass maps.

\begin{figure}
\epsscale{0.6} 
\plotone{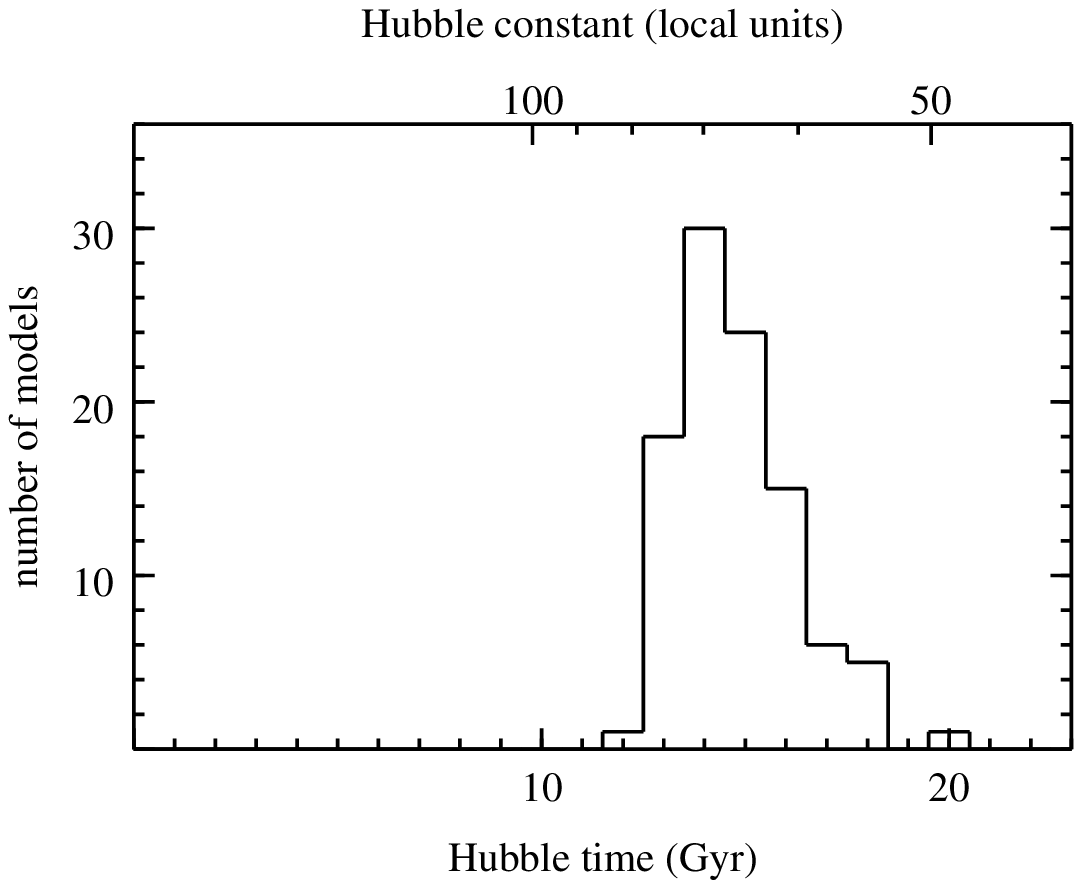}
\caption{Histogram of the Hubble time from 100 four-lens models
of 1520+530, 1600+434, 1830-211, and 2149-275.}
\label{nh2}
\end{figure}

Figure \ref{mass2} shows the common histogram of Hubble times, from
which we obtain
$$\matrix{
  {H_0}^{-1}=14.5_{-1.1}^{+1.6} \rm\ Gyr
& (H_0=67_{-7}^{+6} \rm\ local\ units)
& \hbox{at 68\% confidence} \cr
  {H_0}^{-1}=14.5_{-1.5}^{+3.3} \rm\ Gyr
& (H_0=67_{-13}^{\,+\;8} \rm\ local\ units)
& \hbox{at 90\% confidence}
}$$

\begin{figure}
\epsscale{0.47}
\plotone{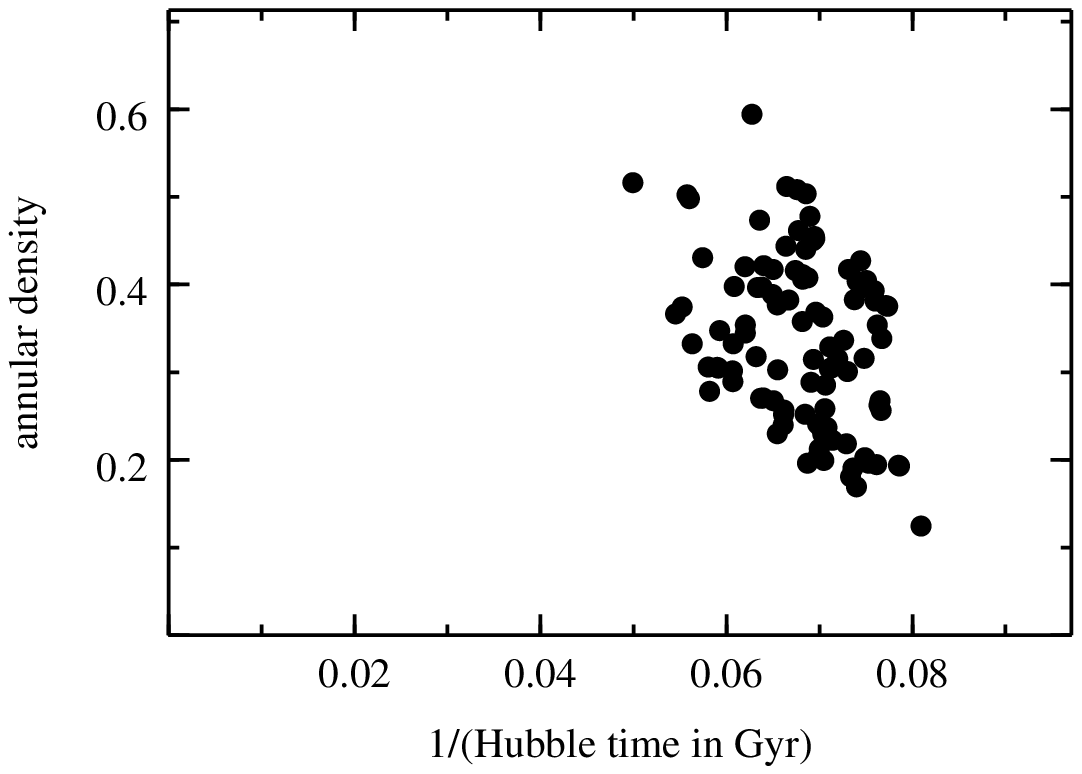}\plotone{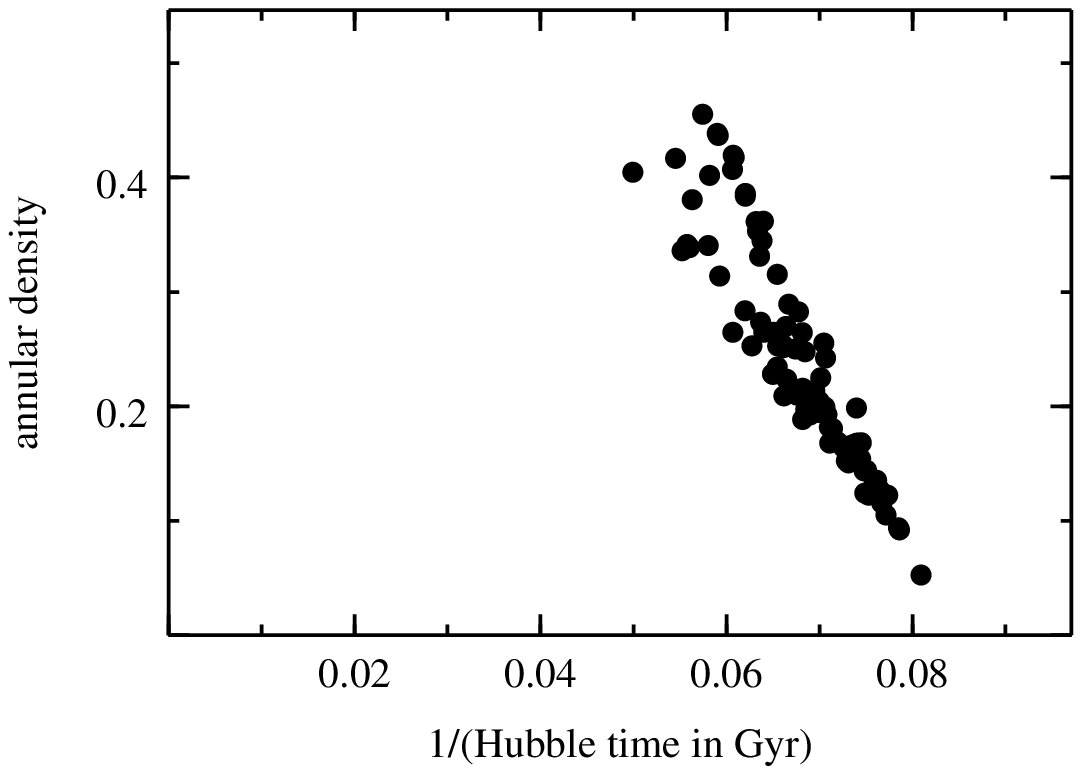}\par
\plotone{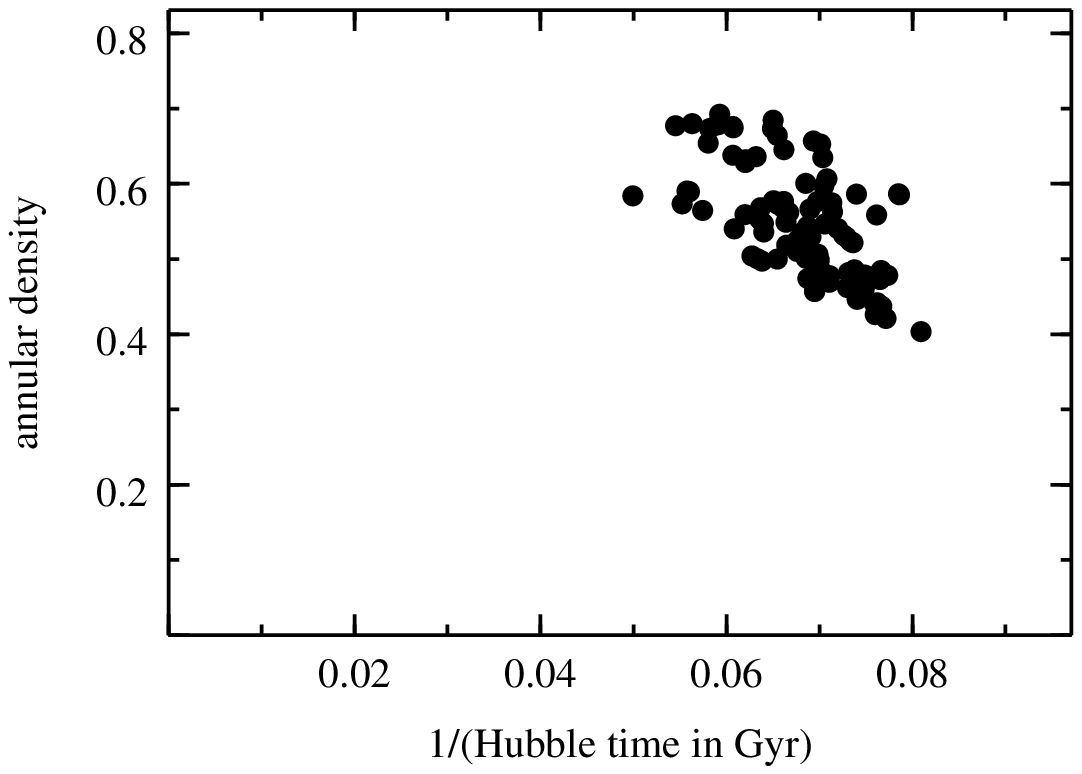}\plotone{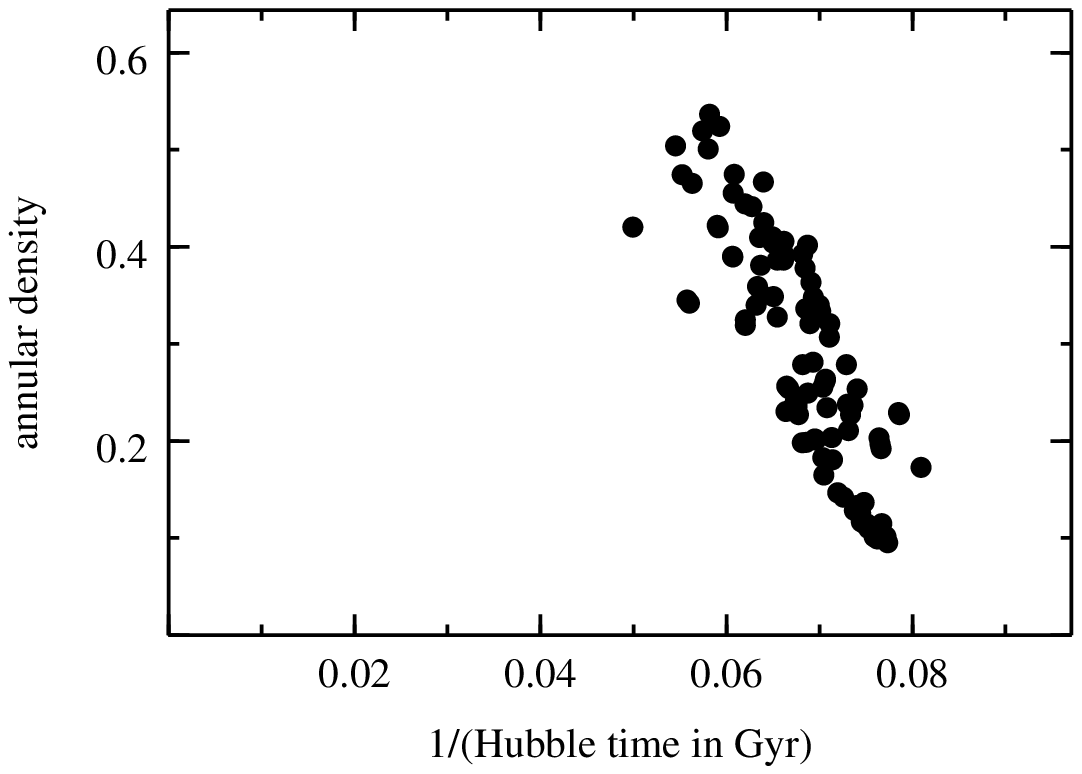}
\caption{Annular density $\kann$ as in Figure \ref{kann}, but for
1520+530 (upper left), 1600+434 (upper right), 1830-211 (lower left),
and 2149-275 (lower right).}
\label{kann2}
\end{figure}

Figure~\ref{kann2} shows $\kann$ against $g^{-1}$ for the four
doubles.  All four show a correlation, 1600+434 being the best.
These correlations are somewhat worse compared to what they would have
looked like if the lens systems were modeled independently (as the
reader can readily verify using {\em PixeLens\/}): as explained in
Section ~\ref{stats}, coupling lens systems weakens the correlations.

\section{Computational issues: reconciling portability and efficiency}

The first decision we made about {\em PixeLens\/} was to write it in
Java. The overriding consideration was portability: we wanted to code
to run without modification on a high-end workstation, a laptop
computer used during a talk, or inside a web browser.  While
developing the software we also found other advantages of Java.  For
example, the graphics libraries are part of the standard, making GUIs
very easy to write.  Also, the language is remarkably
clean\footnote{Apart from a culture of capital letters inside words,
which made the name {\em PixeLens\/} inevitable.}  and easy to debug.

But these advantages come at a price, and we found there were four
issues we had to confront.

First, Java still lacks compilers optimized for numerical work.  Java
numerical code runs slower than C++ by a factor of 3 or more, and also
takes up more memory.  This is an annoying inconvenience, but it is
not fatal.  (Given Moore's law, it is like running C++ on a
two-year-old computer.)

Second, there is very little support for scientific programming.  In
particular, in all the elaborate Java graphics libraries, there is no
simple package to do a plot of $x$ against $y$!  In response, we just
wrote whatever utility software we needed. For example, we wrote Java
code to generate PostScript code to draw the big and small ticks in an
$x,y$ plot.  But this utility code is now available for other projects
and to other researchers.

Third, disk input/output is not allowed from inside a web browser.
This is an essential security precaution---clearly, untrusted code
running over the internet must not be allowed disk access.
Fortunately, the language allows an elegant workaround. {\em PixeLens\/} 
checks whether it is running inside a web browser or as a
standalone program (as an applet or an application in the jargon) and
activates disk input/output only in the latter case.  So users can try
the code out over the internet inside a web browser; if they are
interested enough to want disk output, they can download and run like
a normal program.

Fourth, and most importantly, there is no batch mode inside a web
browser.  If the program takes more than a few seconds to produce
results, which user will want to wait in front of their web browser?
This seems to us the fundamental reason why numerically-intensive Java
programs were previously unknown.  One can write such programs and run
them in batch mode in the normal way, but if they are useless inside a
web browser the single largest advantage of Java is lost.  But there
is a way out.  That way is to provide the user with intermediate
results, and the means to post-process them in non-trivial ways, while
the main computation is still running.  With {\em PixeLens,} the user
can start examining the results as soon as the first model of the
ensemble is complete, even though only 1\% of the total computation
may have run.  We found that providing the user with intermediate
results and sufficient flexibility to make them interesting influenced
almost every aspect of our program design.

\begin{figure}
\epsscale{0.7}
\plotone{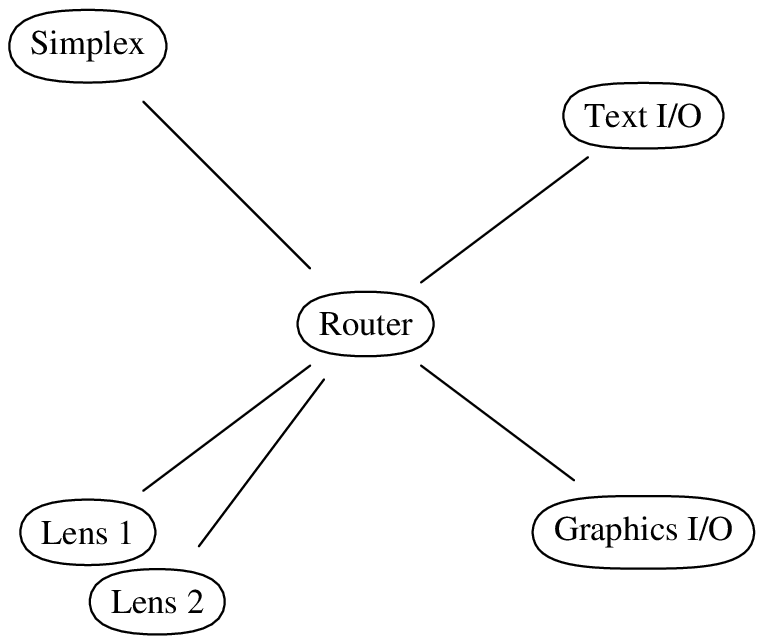}
\caption{Schematic representation of how the code in {\em PixeLens\/}
is organized.}
\label{schem}
\end{figure}

{\em PixeLens\/} is internally documented in a literate programming
idiom \citep{knuth} and we will not enter into minutae here, but
Figure \ref{schem} gives a general idea of how the code is organized.
Imagine each of the labelled ovals as consisting of many variables and
several functions that can operate on those variables.  Functions from
one oval may know something about the variables and functions of
another oval, but only the bare minimum needed to work together.

(Text I/O) is the simplest of the ovals.  It accepts the data and
other input like `run' and `pause', and outputs diagnostics and error
messages, but does not interpret the input itself.  (Router)
interprets enough of the input to work out how many lenses there are,
allocates the required number of (Lens) ovals---two in Figure
\ref{schem}---and passes the data to them.  The (Lens) ovals are the
only parts that know about lensing theory and pixelation.  Each of
them turns the data into constraint equations.  (Router) then takes
the constraint equations from each (Lens), packs them into one large
set of equations, and gives them to (Simplex).  (Simplex) generates
solutions using linear programming and Monte-Carlo, and it does nearly
all the numerical work involved.  Each time a new solution is ready,
(Simplex) passes it to (Router), which unpacks it and distributes
among the (Lens) ovals.  Each (Lens) oval then post-processes the
solution, preparing grids of $\kappa$ and $\tau\(\btheta\)$ and so on.
The post-processed material is passed via (Router) to (Graphics I/O).
(Graphics I/O) can interact with the user and do some non-trivial
computation on its own: it can make contour plots, scatter plots, and
histograms to the user's specification, all without needing to know
any lensing theory.  Moreover (Graphics I/O) can work without
interrupting (Simplex) or the other ovals, even for error-recovery. In
effect, model generation happens in the background while the user
examines intermediate results in the foreground.

\section{Conclusions}

This paper continues our work on modeling lens quasars using pixelated
mass maps.  We elaborate on ideas introduced in earlier papers
\citep{sw97,ws00,sw01,rsw03}: (a)~formulating lens reconstruction as
an undetermined linear inverse problem, (b)~searching through model
space to infer $H_0$ (or alternatively, predict time delays) together
with systematic uncertainties, (c)~predicting the main features of any
Einstein ring, and if observed, estimating the size of the host
galaxy.  But more importantly, we introduce and implement two new
ideas: (i)~reconstructing several time-delay lenses simultaneously
while requiring them to agree about $H_0$, (ii)~making the software so
portable that it can be run inside a web browser while reading this
paper.

The results in this paper are from two worked examples of seven lenses
in all.
\begin{enumerate}
\item First we model the time-delay quads 1115+080 and 1608+656, and
the `dec' 1933+503.  The two time-delay lenses help check against our
earlier work using a different code. By coupling them with 1933+503 we
can predict time-delays for the latter that are not conditional on
$H_0$.  We also obtain
$$ \matrix{
  {H_0}^{-1}=14.6_{-1.7}^{+9.4} \rm\ Gyr
& (H_0=67_{-26}^{\,+\;9} \rm\ local\ units)
& \hbox{at 90\% confidence.} \cr
}$$
\item Then we model the time-delay doubles
1520+530, 1600+434, 1830-211, and 2149-275, and here the main result
$$\matrix{
  {H_0}^{-1}=14.5_{-1.5}^{+3.3} \rm\ Gyr
& (H_0=67_{-13}^{\,+\;8} \rm\ local\ units)
& \hbox{at 90\% confidence}
}$$
\end{enumerate}
The reference cosmology has $(\Omega_M=0.3,\Omega_\Lambda=0.7)$.
Note that the two worked examples contain no lenses 
in common, so the close similarity of the two derived
values of $H_0$ (1 and 2 above) is very encouraging.

The well-known correlations---(i)~steeper lenses give higher $H_0$,
and (ii)~more mass in the annulus covering the images gives lower
$H_0$---are present for most of the lenses, and of these (ii) is
better. 
However, such simple correlations appear to get weaker as $H_0$ is 
better constrained, which happens as a consequence of coupling 
lenses.\footnote{A famous comment by R.O. Redman [quoted by
\cite{gmbp}] comes to mind: hearing ``after all, a star is a pretty
simple thing,'' Redman retorted ``at a distance of 10 parsecs {\it
you'd\/} look pretty simple''.}
Furthermore, lenses with no time delays, like 1933+503, as well as
lenses with very asymmetric mass distributions, like 1608+656, are not 
expected to show the above correlations.

At this stage we remain somewhat cautious about the Hubble-time
estimates.  As we explained in subsection~\ref{prior-sec}, the
distribution of models we obtain is conditional on the prior we use.
In the past, when the uncertainties on the Hubble time were very
large, fine-tuning the prior was not so important.  But now, with the
uncertainties shrinking down from better data and improved modeling,
improving the prior is probably the next priority.

\appendix

\section{Units}

In lensing the Hubble time $H_0^{-1}$ plays a more fundamental role
than the Hubble constant $H_0$.  Note that the crucial observable has
the same dimensions as $H_0^{-1}$ even though its value is many orders
of magnitude different.

Accordingly, rather than using the traditional dimensionless number
$h$ we adopt a dimensionless number $g$, which is defined by
\begin{equation}
  H_0^{-1} = g \rm\; Gyr.
\end{equation}
Numerically, $g=9.78/h$.

To obtain convenient units for the computations, we define
\begin{equation}
\Tzls \equiv (1+z_{\rm L}) {D_{\rm L}D_{\rm S}\over cD_{\rm LS}}
\end{equation}
where the distances are computed using a fiducial value
$H_0^{-1}=1\rm\,Gyr$ (and in some chosen cosmological model).
Multiplying further by $3.6525\times10^{11}/206265^2$ converts $\Tzls$
to units of $\rm days\;arcsec^{-2}$.  Then, by using $\Tzls$ in
appropriate places we can leave all positions in arcsec and all time
delays in days.

We remark that in currently-favored cosmological models, $H_0^{-1}$ is
nearly equal to age of the universe.  In a flat cosmology containing
only matter and dark energy, the Hubble time is related to the age by
(cf.~equation 5.63 in \cite{peeblesbook}.
\begin{equation}
H_0\times\<age> = 
\int_0^1 \left(a\over\Omega_M+(1-\Omega_M)a^3\right)^{1/2} \, da .
\end{equation}
If $\Omega_M=0.3$ then $H_0^{-1}=0.965\,\<age>$. For
$\Omega_M\simeq0.262$ the Hubble time would equal the age; in other
words, the current expansion rate would equal the historic mean.

\acknowledgements


\newpage

\begin{thebibliography}{99}

\bibitem[AbdelSalam et al.(1998)]{asw98}
AbdelSalam, H.M., Saha, P., \& Williams, L.L.R., 1998, \aj, 116, 1541

\bibitem[Barkana(1997)]{barkana97}
Barkana, R, 1997, \apj, 489, 21

\bibitem[Biggs et al.(1999)]{biggs99}
Biggs, A. D., Browne, I. W. A., Helbig, P., Koopmans, L. V. E., Wilkinson, P. N., 
Perley, R. A. 1999, \mnras, 304, 349

\bibitem[Binney et al.(1991)]{binney91}
Binney, J., Gerhard, O. E., Stark, A. A., Bally, J. \& Uchida, K. I.
1991, \mnras, 252, 210

\bibitem[Burud et al.(2000)]{burud00}
Burud, I. et al.\ 2000, \apj, 544, 117

\bibitem[Burud et al.(2002a)]{burud02a}
Burud, I. et al.\ 2002, A\&A, 383, 71

\bibitem[Burud et al.(2002b)]{burud02b}
Burud, I. et al.\ 2002, A\&A, 391, 481

\bibitem[Chavushyan et al.(1997)]{chavushyan97}
Chavushyan, V.H., Vlasyuk, V.V., Stepanian, J.A., Erastova, L.K. 1997,
A\&A, 318, L67

\bibitem[Cohen et al.(2000)]{cohen00}
Cohen, A.S., Hewitt, J.N., Moore, C.B., Haarsma, D.B. 2000
\apj, 545, 578

\bibitem[Courbin et a.(2002)]{courbin02}
Courbin, F., Meylan, G., Kneib, J.-P. \& Lidman, C. 2002, \apj, 575, 95

\bibitem[Fassnacht et al.(2002)]{fassnacht02}
Fassnacht, C.D., Xanthopoulos, E., Koopmans L.V.E., Rusin D. 2002,
\apj, 581, 823

\bibitem[Gerhard et al.(2001)]{gerhard01}
Gerhard, O., Kronawitter, A., Saglia, R. P., Bender, R.
2001, \aj, 121, 1936

\bibitem[Heggie \& Hut(2003)]{gmbp}
Heggie, D.C. \& Hut, P. 2003,
{\em The Gravitational Million-Body Problem,}
Cambridge University Press.

\bibitem[Hjorth et al.(2002)]{hjorth02}
Hjorth, J. et al.\ 2002, \apj, 572, L11

\bibitem[Impey et al.(1998)]{impey98}
Impey, C.D., Falco, E.E., Kochanek, C.S., Leh\'ar, J., McLeod, B.A.,
Rix, H.-W., Peng, C.Y., \& Keeton, C.R. 1998, \apj, 509, 551

\bibitem[Jackson et al(1995)]{jackson95}
Jackson, N., et al.\ 1995, MNRAS, 274, L25

\bibitem[Keeton(2001)]{keeton01}
Keeton, C.R. 2001, {\em gravlens,}
available at {\tt cfa-www.harvard.edu/glensdata}

\bibitem[Keeton \& Winn(2003)]{kw03}
Keeton, C.R., \& Winn, J.N. 2003, \apj, 590, 39

\bibitem[Kochanek et al.(1998)]{castles}
Kochanek, C.S., Falco, E.E., Impey, C., Leh\'ar, J., McLeod, B., Rix,
H.-W. 1998,\\
{\tt cfa-www.harvard.edu/glensdata}

\bibitem[Kochanek(2002)]{kochanek02}
Kochanek, C.S. 2002, \apj, 578, 25

\bibitem[Koopmans \& de Bruyn(2003)]{kdb03}
Koopmans, L.V.E. \& de Bruyn, A.G. 2003, preprint, astro-ph/0311567

\bibitem[Knuth(1992)]{knuth}
Knuth, D.E. 1992, {\em Literate Programming,} CSLI publications.

\bibitem[Lehar et al.(2000)]{lehar00}
Lehar, J. et al.\ 2000, \apj, 536, 584

\bibitem[Lovell et al.(1998)]{lovell98}
Lovell, J.E.J., Jauncey, D.L., Reynolds, J.E., Wieringa, M.H., King,
E.A., Tzioumis, A.K., McCulloch, P.M., Edwards, P.G., 1998,
\apj 508, L51

\bibitem[Myers et al.(1995)]{myers95}
Myers, S.T. et al.\ 1995, \apjl, 447, 5

\bibitem[Oscoz et al.(2001)]{oscoz01}
Oscoz, A., et al.\ 2001, \apj, 552, 81

\bibitem[Peebles(1993)]{peeblesbook}
Peebles, P.J.E. 1993, {\em Principles of Physical Cosmology}, Princeton

\bibitem[Rao \& Subrahmanyan(1988)]{rao88}
Rao, A., \& Subrahmanyan, R. 1988, MNRAS, 231, 229

\bibitem[Raychaudhury et al.(2003)]{rsw03}
Raychaudhury, S., Saha, P. \& Williams, L.L.R. 2003, \aj, 126, 29

\bibitem[Rix et al.(1997)]{rix97}
Rix, H.-W., de Zeeuw, P. T., Cretton, N., van der Marel, R. P., Carollo, C. M.
1997, \apj, 488, 702

\bibitem[Saha(2003)]{pda}
Saha, P. 2003, {\em Principles of Data Analysis,} Cappella Archive.

\bibitem[Saha \& Williams(1997)]{sw97}
Saha, P., \& Williams, L.L.R. 1997, \mnras, 292, 148

\bibitem[Saha \& Williams(2001)]{sw01}
Saha, P., \& Williams, L.L.R. 2001, \aj, 122, 585

\bibitem[Saha \& Williams(2003)]{sw03}
Saha, P., \& Williams, L.L.R. 2003, \aj, 125, 2769

\bibitem[Schechter et al.(1997)]{schechter97}
Schechter P.L. et al.\ 1997, \apj, 475, L85

\bibitem[Sykes et al.(1998)]{sykes98}
Sykes, C.M. et al.\ 1998, \mnras, 301, 310

\bibitem[Trotter et al.(2000)]{trotter00}
Trotter, C.S., Winn, J.N., Hewitt, J.N. 2000, \apj, 535, 671

\bibitem[Young et al.(1981)]{young81}
Young, P., Gunn, J.E., Oke, J.B., Westphal, J.A., Kristian, J. 1981,
\apj, 244, 736

\bibitem[Weymann et al.(1980)]{weymann80}
Weymann, R.J. 1980, Nature, 285, 641

\bibitem[Williams \& Saha(2000)]{ws00}
Williams, L.L.R. \& Saha, P. 2000, \aj, 119, 439

\bibitem[Winn et al.(2002)]{winn02}
Winn, J.N., Kochanek, C.S., McLeod, B.A., Falco, E.E., Impey, C.D. \& Rix, H.-W.
2002, \apj, 575, 103

\bibitem[Wisotzki et at(1996)]{wisotzki96}
Wisotzki, L., Koehler, T., Lopez, S., \& Reimers, D. 1996,
A\&A, 315, L405 

\end{thebibliography}
\end{document}